\documentclass[prd,twocolumn,superscriptaddress,nofootinbib,noshowpacs,preprintnumbers,longbibliography,floatfix]{revtex4-2}
\usepackage[utf8]{inputenc}
\usepackage{graphicx}
\usepackage{placeins,float}
\usepackage{amssymb}
\usepackage{amsmath}  
\usepackage{mathtools}
\usepackage{dsfont}
\usepackage{array}
\usepackage{bm,fixmath}
\usepackage{mathrsfs}
\usepackage{pifont}
\usepackage{multirow}
\usepackage{upgreek}
\usepackage{overpic}
\usepackage{xcolor}
\usepackage{bm}
\usepackage{bbm}
\usepackage{physics}
\usepackage{comment}
\usepackage{tikz}
\usetikzlibrary{quantikz}
\usepackage{subfigure}
\usepackage[pdftex,
            pdftitle={First-Order Phase Transition of the Schwinger Model with a Quantum Computer},
            pdfauthor={Takis Angelides, Pranay Naredi, Arianna Crippa, Karl Jansen, Stefan Kuehn, Ivano Tavernelli, Derek S. Wang},
            bookmarks,
            colorlinks,
            linkcolor=myblue,
            citecolor=mymagenta,
            menucolor=black,
            urlcolor=myblue,
            plainpages=false,
            pdfpagelabels,
            hypertexnames=false]{hyperref}
\graphicspath{{figures/}}
\definecolor{mymagenta}{RGB}{200, 0, 100}
\definecolor{myblue}{RGB}{45, 48, 146}
\definecolor{mygreen}{RGB}{0, 126, 0}
\definecolor{myorange}{RGB}{255, 136, 19}
\usepackage{orcidlink}

\begin{document}

\title{First-Order Phase Transition of the Schwinger Model with a Quantum Computer}

\author{Takis Angelides \orcidlink{0000-0002-8639-8050}}
\affiliation{Institut für Physik, Humboldt-Universität zu Berlin, Newtonstr. 15, 12489 Berlin, Germany}
\affiliation{CQTA, Deutsches Elektronen-Synchrotron DESY, Platanenallee 6, 15738 Zeuthen, Germany}

\author{Pranay Naredi \orcidlink{0000-0002-4357-6718}}
\affiliation{Computation-Based Science and Technology Research Center, The Cyprus Institute, 20 Kavafi Street,
2121 Nicosia, Cyprus}

\author{Arianna Crippa \orcidlink{0000-0003-2376-5682}}
\affiliation{Institut für Physik, Humboldt-Universität zu Berlin, Newtonstr. 15, 12489 Berlin, Germany}
\affiliation{CQTA, Deutsches Elektronen-Synchrotron DESY, Platanenallee 6, 15738 Zeuthen, Germany}

\author{Karl Jansen \orcidlink{0000-0002-1574-7591}}
\affiliation{CQTA, Deutsches Elektronen-Synchrotron DESY, Platanenallee 6, 15738 Zeuthen, Germany}
\affiliation{Computation-Based Science and Technology Research Center, The Cyprus Institute, 20 Kavafi Street,
2121 Nicosia, Cyprus}

\author{Stefan K\"uhn \orcidlink{0000-0001-7693-350X}}
\affiliation{CQTA, Deutsches Elektronen-Synchrotron DESY, Platanenallee 6, 15738 Zeuthen, Germany}

\author{Ivano Tavernelli \orcidlink{0000-0001-5690-1981}}
\affiliation{IBM Research Europe - Zurich, Säumerstrasse 4, 8803 Rüschlikon, Switzerland}

\author{Derek S.\ Wang \orcidlink{0000-0003-4538-5816}}
\affiliation{IBM Quantum, IBM T.J.\ Watson Research Center, Yorktown Heights, NY 10598, USA}

\date{\today}

\begin{abstract}
    We explore the first-order phase transition in the lattice Schwinger model in the presence of a topological $\theta$-term by means of the variational quantum eigensolver (VQE). Using two different fermion discretizations, Wilson and staggered fermions, we develop parametric ansatz circuits suitable for both discretizations, and compare their performance by simulating classically an ideal VQE optimization in the absence of noise. The states obtained by the classical simulation are then prepared on the IBM's superconducting quantum hardware. Applying state-of-the art error-mitigation methods, we show that the electric field density and particle number, observables which reveal the phase structure of the model, can be reliably obtained from the quantum hardware. To investigate the minimum system sizes required for a continuum extrapolation, we study the continuum limit using matrix product states, and compare our results to continuum mass perturbation theory. We demonstrate that taking the additive mass renormalization into account is vital for enhancing the precision that can be obtained with smaller system sizes. Furthermore, for the observables we investigate we observe universality, and both fermion discretizations produce the same continuum limit.
\end{abstract}

\maketitle
 
\section{Introduction}

Exploring the phase diagram of a theory is central for the understanding of the fundamental laws of physics in many disciplines ranging from condensed matter to particle physics. For instance, in the case of ferromagnetic materials, critical points and phase boundaries help to understand and design magnetic materials with various applications~\cite{Zhang_2021, XIA2023171112}. 
Within superconductivity, phase diagrams allow for identifying the critical temperatures and magnetic fields required for materials to exhibit zero electrical resistance~\cite{text_mine_sc}. In particle physics, the phase diagram of Quantum Chromodynamics (QCD) elucidates the behavior of matter under extreme conditions, relevant for the early universe and neutron stars~\cite{sym15020541}.
In high-energy physics, topological terms play an important role in the study of interesting phenomena such as the breaking of charge conjugation-parity symmetry in QCD~\cite{PhysRevD.19.1826}, and the occurrence of out-of-equilibrium dynamical effects involving axion fields~\cite{PhysRevLett.40.223}.

In order to explore the phase structure of a given theory, one often has to resort to numerical methods. A standard tool is Monte Carlo (MC) simulation, which has been successful for probing phase diagrams~\cite{mc_design, acharyya, berthier, mc_pt}, including that of lattice QCD~\cite{10.1143/PTPS.174.206}. However, the sign problem poses a barrier in certain parameter regimes~\cite{deForcrand:2009zkb}, leaving relevant questions unanswered. Specifically for lattice QCD, large baryon chemical potentials or the presence of a topological $\theta$-term would trigger this problem~\cite{finite_density_lqcd_1}. Thus, variations of the conventional MC approach have been proposed to tackle this obstacle~\cite{10.1143/PTPS.174.206}, but so far only with limited success. Hence, there is great interest in alternative methods that bypass the issue, such as tensor networks and quantum computing~\cite{2023_qc_review_for_lft,Banuls2019,Banuls2020,halimeh2023coldatom}. In particular, quantum computing offers a promising alternative route~\cite{Banuls2019,Banuls2020,DiMeglio2023}, with already a number of successful demonstrations~\cite{2023arXiv230804481F, martinez_2016, PRXQuantum.3.020324, PRXQuantum.2.030334, Bassman_2021, Thompson_2022, jian-wei-et-al, zhang2023observation}. 

Here, we focus on exploring the possibility of studying the phase structure of the Schwinger model~\cite{Schwinger1962} in the presence of a topological $\theta$ term with near-term quantum devices. Despite its simplicity, the Schwinger model in the presence of a topological $\theta$-term provides an example where conventional Monte Carlo methods would suffer from the sign problem. Moreover, it exhibits a rich phase structure with a first-order quantum phase transition for large enough masses at $\theta=\pi$~\cite{angelides, topo_vac, Azcoiti, hamer82, COLEMAN1976239, COLEMAN1975267, PhysRevD.105.014504, Pederiva:2021tcd, byrnes02}. Hence, Schwinger model serves as a benchmark system for developing and testing new methods.  Regarding a quantum computing approach to the model in the presence of a topological $\theta$-term, there are several open questions. First, the theory needs to be discretized on a lattice with a finite extent, and there are various different ways for discretizing fermions on such a lattice. It is therefore a priori not clear which one will show the best performance for a given set of resources. Second, it is important to identify how large the system sizes should be in order to obtain a reliable continuum extrapolation in order to assess the applicability of near-term quantum devices. 

Here we address these questions by studying two fermion discretizations, staggered and Wilson fermions, in order to explore any potential advantages of one type of fermion formulation over the other. We use the variational quantum eigensolver (VQE)~\cite{vqe} and derive a protocol for mapping this problem on a quantum circuit. The VQE, introduced as an alternative to quantum phase estimation~\cite{kitaev}, aligns with the capabilities of current and near-term quantum devices. We test our VQE using noiseless classical simulations between 6 to 12 qubits, to identify the best possible setup regarding the ansatz and gates that would capture the relevant ground states most efficiently. After the optimal ansatz-gate combination and variational parameters are found, the ground states across the phase transition are prepared on IBM's quantum devices. We demonstrate that using state-of-the-art error mitigation techniques---zero noise extrapolation~\cite{ZNE}, readout error mitigation~\cite{trex}, Pauli twirling~\cite{pauli_twirling_for_coherent_noise} and dynamical decoupling~\cite{dynamical_decoupling}---allows for obtaining precise results from the quantum measurements. To understand the minimum system sizes required to extrapolate faithfully to the continuum limit with a quantum computer, we use matrix product states (MPS). We numerically simulate intermediate system sizes and perform the continuum extrapolation, which we compare to analytical results. Our study also shows universality for the considered observables, as both discretizations  lead to the same continuum values.

The rest of the paper is structured as follows. In Sec.~\ref{theory}, we briefly introduce the Schwinger model and review its phase structure in the presence of a topological $\theta$-term. Moreover, we discuss two different ways of discretizing it on a lattice using Wilson and staggered fermions. We proceed with presenting our ansatz for the VQE as well as the MPS techniques we use to estimate the resources for taking a reliable continuum limit in Sec.~\ref{methods}. Our numerical results demonstrating the performance of the ansatz in various parameter regimes are presented in Sec.~\ref{results}, before concluding in Sec.~\ref{sec:conclusion}.

\section{The Schwinger model\label{theory}}

The Schwinger model describes quantum electrodynamics in (1+1)-dimensions coupled to a single, massive Dirac fermion~\cite{Schwinger1962}. Here we briefly introduce the Hamiltonian formulation and review its phase diagram in the presence of a topological $\theta$-term. We then discuss two different discretizations for the fermionic matter fields of the model, namely Wilson and staggered fermions.

\subsection{Hamiltonian formulation in the continuum\label{Continuum_Schwinger_Model}}
The continuum Hamiltonian density of the Schwinger model in the presence of a topological $\theta$-term is given by
\begin{equation}
\label{hamiltonian_density}
    \mathcal{H} = -i\overline{\psi}\gamma^1\left(\partial_1 - igA_1\right)\psi + m\overline{\psi}\psi + \frac{1}{2}\left(\dot{A}_1 + \frac{g\theta}{2\pi}\right)^2,
\end{equation}
where $\psi(x)$ is a two-component Dirac spinor describing the fermionic matter. The spinor components, $\psi_\alpha$, $\alpha=1,2$, fulfill the standard fermionic anticommutation relations $\{\psi_\alpha^\dagger(x),\psi_\beta(y)\} = \delta(x-y)\delta_{\alpha\beta}$. The gauge field $A_\mu$, $\mu=0,1$, mediates the interaction between the matter fields. Here we have chosen the temporal gauge, $A_0=0$, hence only the spatial component $A_1$ appears in the Hamiltonian. The parameters $m$ and $g$ are the bare fermion mass and the coupling between fermions and the gauge fields. The matrices $\gamma^\mu$ are two dimensional matrices obeying the Clifford algebra $\{\gamma^\mu, \gamma^\nu\} = 2\eta^{\mu \nu}$, with $\eta = \text{diag}(1,-1)$, and $\overline{\psi}$ corresponds to $\psi^\dagger \gamma^0$. The physically relevant gauge invariant states of the Hamiltonian have to fulfill Gauss's law
\begin{align}
    -\partial_1\dot{A}^1 = g\overline{\psi}\gamma^0\psi,
    \label{eq:gauss_law}
\end{align}
where $-\dot{A}^1$ is the electric field and $g\overline{\psi}\gamma^0\psi$ represents the charge density. 

\begin{figure}[htp!]
    \centering
    \includegraphics[width=0.9\linewidth]{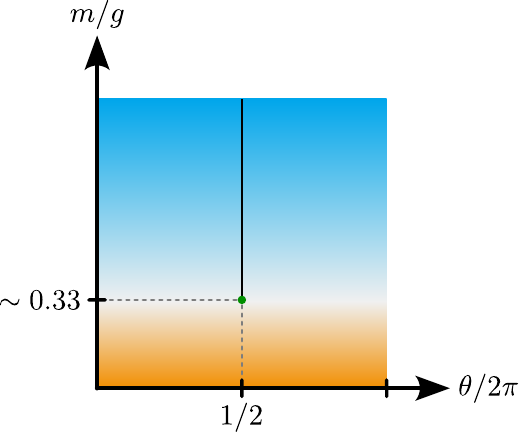}
    \caption{Illustration of the phase diagram of the Schwinger model in the presence of a topological term in the $m/g-\theta$ plane. Since the physics is periodic in $\theta$ with period $2\pi$, only the first period is shown. The critical line (shown in black) indicates the first-order phase transitions occurring at $\theta=\pi$ for masses larger than the critical one $m_c/g\approx 0.33$, which ends in a second-order phase transition (green dot) exactly at $m_c/g$. Below the critical mass no transitions occur.}
    \label{phase_diagram}
\end{figure}

The topological term, $g\theta/2\pi$, appearing in the Hamiltonian corresponds to a constant background electric field whose effect has been assessed both theoretically and numerically. Coleman argued that the physics of the model is periodic in $\theta$ with a period of $2\pi$, and that above a certain critical mass, $m_c/g$, the model undergoes a first order quantum phase transition at $\theta=\pi$~\cite{COLEMAN1976239}. This picture was later on confirmed in numerical simulations, where it was found that the critical line ends in a second-order quantum phase transition at $m_c/g\sim 0.33$~\cite{hamer82,byrnes02,buyens}. Figure~\ref{phase_diagram} provides a sketch of the phase diagram, highlighting the first-order phase transition line, which culminates with a second-order phase transition at $m_c/g$. 
The physics of the model can also be understood qualitatively in an intuitive picture. For large values of the mass in units of the coupling, $m/g\gg 1$, it is generally unfavorable to generate charged particles. In the regime $l_0 = \theta/2\pi < 1/2$, the ground state thus remains devoid of charged particles and the electric field matches the background electric field due to Gauss's law. For $l_0 > 1/2$, it becomes energetically favorable to produce a negative unit charge on the left of the one-dimensional space and a particle of positive unit charge on the right. As a result of Gauss's law, these are connected by a flux string, lowering the electric field by one unit, which takes the electric field in the bulk from $l_0$ to $l_0 - 1$~\cite{byrnes02,COLEMAN1976239}. Given that the electric field energy is proportional to the square of the electric field and the mass contribution of the particle-antiparticle pair is negligible at infinite volume, these two states are degenerate directly at the point $l_0 = 1/2$. Hence, a first-order quantum phase transition occurs at $l_0 = 1/2$ (or equivalently $\theta = \pi$) for large values of $m/g$, manifesting itself in an abrupt change of the electric field density and particle number. This transition spontaneously breaks the invariance under a combined charge conjugation and parity symmetry transformation (CP transformation)~\cite{buyens}.

For small values of $m/g$ below the critical mass, it is energetically favorable to screen the electric field by producing more of the negative-positive charge pairs, and at $m/g = 0$ we have complete screening of the background electric field~\cite{adam}. Hence, in this regime the electric field density and the particle number do not display an abrupt change as a function of $\theta$.

\subsection{Lattice discretizations\label{lattice_discretizations}}

In order to address the Hamiltonian of the theory numerically, we choose to work with a discrete lattice formulation. To ensure that one faithfully recovers the physics of the continuum model in the limit of vanishing lattice spacing, special care has to be taken when discretizing the fermionic degrees of freedom. In particular, a naive discretization of the fermionic fields leads to an incorrect continuum limit due to the so-called doubling problem~\cite{Rothe2006}. There are several ways to avoid this problems, and here we choose to work with two types of fermion discretizations, Wilson and staggered fermions, which are commonly used in the literature. We first turn to Wilson fermions and present the final Hamiltonian as well as the relevant observables in this formulation before discussing the same operators for the staggered discretization. Moreover, we show how to translate them to qubit degrees of freedom.

\subsubsection{Wilson fermions\label{wilson_fermions_subsection}}

The basic idea of Wilson fermions is to add a lattice version of the second derivative of the fermion field to the native discretization (for details, see Appendix~\ref{wilson_operators}). While this term vanishes in the continuum limit, it can be shown that it breaks the chiral symmetry explicitly, and gives the doubler solutions originating from the discretization of space a mass proportional to the inverse of the lattice spacing. This ensures that they decouple from the theory when approaching the continuum limit~\cite{wilson_fermions_original_work, Creutz1994}. The Wilson discretization of the of the Schwinger Hamiltonian on a lattice with $N$ sites and spacing $a$ reads~\cite{angelides, Zache}
\begin{align}
    \label{interacting_discrete_hamiltonian}
    \begin{aligned}
        H_W &= \sum_{n=0}^{N-2}\left(\overline{\phi}_n\left(\frac{r+i\gamma^1}{2a}\right)U_n\phi_{n+1} + \text{h.c.} \right) \\
        &+\sum_{n=0}^{N-1} \left(m_\text{lat}+\frac{r}{a}\right)\overline{\phi}_n\phi_n + \sum_{n=0}^{N-2}\frac{ag^2}{2}\left(L_n+l_0\right)^2.
    \end{aligned}
\end{align}
The field $\phi_n$ is a dimensionless two-component Dirac spinor at site $n$, whose components $\phi_{n,\alpha}$, $\alpha = 1,2$, again fulfill the standard anticommutation relations for fermionc fields $\{\phi_{n,\alpha}^\dagger,\phi_{n',\beta}\} = \delta_{nn'}\delta_{\alpha\beta}$. The operators $U_n$ and $L_n$ act on the links in between the matter sites $n$ and $n+1$, where $L_n$ represents the quantized dimensionless electric field. They fulfill the commutation relation $[U_n,L_{n'}] = \delta_{nn'}U_n$, hence, in the eigenbasis of the electric field operator, $U_n$ acts as a lowering operator, decreasing the electric flux value by one unit. The parameters $m_{\text{lat}}$ and $g$ are the lattice mass and coupling, and $r$ corresponds the Wilson parameter, which can be chosen arbitrarily between $(0,1]$ to ensure the correct continuum limit~\cite{Zache}. The background electric field, $l_0=\theta/2\pi$, is the lattice version of the topological $\theta$-term. On the lattice, Gauss's law translates to
\begin{align}
    L_n - L_{n-1} = Q_n
    \label{eq:gauss_law_lattice}
\end{align}
where $Q_n = \phi_n^\dagger \phi_n - 1$ is the charge operator. Given the global $U(1)$ symmetry of the model, the total charge $\sum_n Q_n$ is conserved. The lattice discretization with Wilson formulation generates an additive mass renormalization~\cite{angelides}. The renormalized mass can be expressed as
\begin{equation}
    \label{mass_shift_equation}
    m_r/g = m_{\text{lat}}/g + \text{MS}(V,ag,l_0),
\end{equation}
where MS is the additive mass shift which, in general, depends on the dimensionless lattice volume, $V = N/\sqrt{x}$, the lattice spacing in units of the coupling, $ag$, and the background field, $l_0$. In order to follow a line of constant $m_\text{r}/g$, one needs to determine MS. For our simulations later on, we use the method from Ref.~\cite{angelides} to obtain MS. For the Wilson discretization, we can define the particle number operator as
\begin{equation}
    P_W = N + \sum_{n=0}^{N-1} \overline{\phi}_n\phi_n,
\end{equation}
which essentially is the lattice version of the chiral condensate, $\overline{\psi}\psi$, up to the constant $N$. The constant shift ensures that $P_W$ is a positive semidefinite operator, where the zero eigenvalue is obtained for states minimizing $\sum_{n=0}^{N-1} \overline{\phi}_n\phi_n$. These correspond to the ground state of the Hamiltonian in the limit of dominating mass term, for which we do not expect any particles from the picture discussed for the continuum model. As soon as particles are present in the ground state, we expect a nonzero expectation value for $P_W$. Thus, together with the electric field, this observable should allow us to detect the first-order phase transition in the lattice model.

For open boundary conditions, Eq.~\eqref{eq:gauss_law_lattice} can be solved iteratively after fixing the value of the electric field on the left boundary, $\varepsilon_0$. Choosing $\varepsilon_0=0$, we find for the electric field\footnote{Choosing a nonzero $\varepsilon_0$ value for the electric field can be interpreted as a shift in the value of $l_0$ to $l_0' = l_0 + \varepsilon_0$. Hence, we can set $\varepsilon_0=0$ without loss of generality.}
\begin{align}
    L_n = \sum_{k=0}^nQ_k. 
    \label{eq:integrating_gauss_law}
\end{align}
This implies that the electric field on each link can be determined by the fermionic charges, and is not a dynamical variable. Inserting this into the Hamiltonian in Eq.~\eqref{interacting_discrete_hamiltonian} and applying a residual transformation to the fermionic fields~\cite{hamer97}, we can integrate out the gauge fields and obtain a formulation purely in fermionic degrees of freedom~\cite{angelides}
\begin{align}    
    \begin{aligned}
        H_W &= \sum_{n=0}^{N-2}\left(\overline{\phi}_n\left(\frac{r+i\gamma^1}{2a}\right)\phi_{n+1} + \text{h.c.} \right) \\
        &+\sum_{n=0}^{N-1} \left(m_\text{lat}+\frac{r}{a}\right)\overline{\phi}_n\phi_n + \sum_{n=0}^{N-2}\frac{ag^2}{2}\left(l_0+ \sum_{k=0}^{n} Q_k\right)^2.    
    \end{aligned}
    \label{eq:interacting_discrete_hamiltonian_no_gauge_fields}
\end{align}
Note that the above Hamiltonian is directly confined to the gauge invariant subspace, as Gauss's law is fulfilled by construction. For the rest of this paper, we will fix $r=1$ and choose to work with the convention $\gamma^0=X$, $\gamma^1 = iZ$, where $X$, $Z$ are the usual Pauli matrices.

In order to be able to measure the Hamiltonian on a quantum device, we have to map the fermionic degrees of freedom to qubits. To this end, we use the Jordan-Wigner transformation~\cite{jordan_wigner} to translate the fermion fields to spins. Note that in the Wilson formulation, we have a two-component spinor on each site. Hence, the Jordan-Wigner transformation for a system with $N$ sites results in $2N$ spin degrees of freedom. As shown in detail in Appendix~\ref{wilson_operators}, applying a convenient ordering for the fermions, allows for obtaining a purely real Hamiltonian, and we find for the dimensionless spin formulation of Eq.~\eqref{eq:interacting_discrete_hamiltonian_no_gauge_fields}
\begin{align}
\label{wilson_hamiltonian}
    \begin{aligned}
        W_W &= x\sum_{n=0}^{N-2}\left(X_{2n+1}X_{2n+2} + Y_{2n+1}Y_{2n+2}\right) \\
        &+ \left(\frac{m_\text{lat}}{g}\sqrt{x} + x\right)\sum_{n=0}^{N-1}\left(X_{2n}X_{2n+1} + Y_{2n}Y_{2n+1}\right) \\
        &+\sum_{n=0}^{N-2}\left(l_0 + \sum_{k=0}^{n} Q_k\right)^2.
    \end{aligned}
\end{align}
Here we have defined the inverse lattice spacing squared in units of the coupling, $x = 1/(ag)^2$. The charge operator in spin formulation is given by $Q_n = (Z_{2n} + Z_{2n+1})/2$, and $X_n$, $Y_n$, and $Z_n$ are the standard Pauli operators acting on spin $n$.

The relevant observables for detecting the first-order phase transition, can be readily translated to the spin formulation. Since we expect the electric field density to be homogeneous in the bulk for large enough system sizes, we only consider a single link operator in the center of the system to minimize boundary effects. The total electric field including the background field on this link can then be reconstructed from the charges as
\begin{equation}
\label{L_middle}
    \begin{aligned}
        L_W &=l_0 + L_{\lceil\frac{N}{2}\rceil-1} = l_0 + \sum_{k=0}^{\lceil N/2 \rceil-1} Q_k \\
        &= l_0 + \frac{1}{2}\sum_{k=0}^{\lceil N/2 \rceil-1} (Z_{2k} + Z_{2k+1}).
    \end{aligned}
\end{equation}
Using the same ordering for the fermions as for the Hamiltonian, the particle number operator becomes
\begin{equation}
    P_W = N + \frac{1}{2}\sum_{n=0}^{N-1} \left(X_{2n}X_{2n+1} + Y_{2n}Y_{2n+1}\right).
\end{equation}
This operator commutes with the Hamiltonian only in the weak-coupling limit, $m/g \gg 1$, for which the kinetic term can be neglected. Hence, only in this limit we expect the integer eigenvalues of $P_W$ to be good quantum numbers~\cite{PhysRevD.96.114501}. 

\subsubsection{Staggered fermions\label{staggered_fermions_subsection}}

Another approach to discretize the model are the Kogut-Susskind staggered fermions~\cite{KS_hamiltonian_formulation}. The basic idea is to distribute the two spinor components to different lattice sites, thereby ``thinning out'' the fermionic degrees of freedom and avoiding the doubling problem. The Kogut-Susskind formulation for the Schwinger Hamiltonian for a lattice with $N$ sites, spacing $a$ and open boundary conditions can written as~\cite{KS_hamiltonian_formulation}
\begin{align}
\label{Staggered_hamiltonian_initial}
    \begin{aligned}
        H_S &= -\frac{i}{2a}\sum_{n=0}^{N-2} \left(\phi^\dagger_n U_n\phi_{n+1}-\text{h.c}.\right)  \\
        &+ m_\text{lat}\sum_{n=0}^{N-1} (-1)^n\phi^\dagger_n\phi_n + \frac{ag^2}{2}\sum_{n=0}^{N-2} L_n^2 \,,
    \end{aligned}
\end{align}
where $N$ is supposed to be even. In the expression above, the operators $L_n$ and $U_n$ are the same as for the Wilson case and act on the links between sites $n$ and $n+1$. Different to the Wilson discretization, $\phi_n$ now represents a single-component fermionic field whose anticommutation relations are given by $\{\phi_n^\dagger,\phi_{n'}\} = \delta_{nn'}$. In the limit of vanishing lattice spacing, the fermionic fields residing on even (odd) sites will correspond to the upper (lower) component of the Dirac spinor $\psi(x)$. Thus, the staggered formulation effectively doubles the lattice spacing. Again, $m_\text{lat}$ and $g$ refer to the lattice mass and the coupling. The Gauss's law for the staggered case has the same form as in Eq.~\eqref{eq:gauss_law_lattice}, but $Q_n$ is now given by the staggered charge operator $Q_n = \phi^\dagger_n\phi_n-\left(1-(-1)^n\right)/2$. Also for the staggered formulation, the total charge $\sum_n Q_n$ is a conserved quantity and commutes with the Hamiltonian. 

Similar to the Wilson case, it was recently shown that staggered fermions experience an additive mass renormalization, where the renormalized mass can again be written as in Eq.~\eqref{mass_shift_equation}~\cite{chiral_dempsey_staggered}. For the case of periodic boundary conditions, Ref.~\cite{chiral_dempsey_staggered} was able to compute the mass shift analytically finding
\begin{align}
    \text{MS}_t = \frac{1}{8\sqrt{x}}.
    \label{eq:ms_staggered}
\end{align}
For open boundary conditions, there is no analytical prediction, and we can measure the mass shift using the approach from Ref.~\cite{angelides}. Moreover, Ref.~\cite{angelides} also demonstrated that the results for MS obtained for the staggered discretization using open boundary conditions agree with the theoretical prediction $\text{MS}_t$ for large lattice volumes. Thus, for large enough volumes Eq.~\eqref{eq:ms_staggered} should still provide an approximation even for open boundary conditions. In our numerical computations later on, we will consider both options. 

Analogously to the Wilson case, the particle number operator for the staggered fermions resembles the mass term of the Hamiltonian and is given by
\begin{equation}
    P_S = \frac{N}{2} + \sum_{n=0}^{N-1}(-1)^n\phi_n^\dag\phi_n
\end{equation}
This expression is again equivalent the lattice version of the chiral condensate up to a constant shift, and has the same properties as for the Wilson case. Note that $P_S$ now contains an alternating sign factor, which is the result of the staggered discretization separating the spinor components to different lattice sites. Together with the electric field, this observable is again suitable to detect the first-order phase transition in the model. 

As in the case of Wilson fermions, the gauge fields can be integrated using Gauss's law resulting in Eq.~\eqref{eq:integrating_gauss_law}, but now with the staggered charge. Performing the same steps as in the Wilson case, we can obtain a Hamiltonian directly restricted to the gauge invariant subspace where all notion of the gauge field is gone~\cite{hamer97}
\begin{align}
    \begin{aligned}
            H_S &= -\frac{i}{2a}\sum_{n=0}^{N-2} \left(\phi^\dagger_n \phi_{n+1}-\text{h.c}.\right)  \\
            &+ m_\text{lat}\sum_{n=0}^{N-1} (-1)^n\phi^\dagger_n\phi_n + \frac{ag^2}{2}\sum_{n=0}^{N-2}\left(l_0+ \sum_{k=0}^{n} Q_k\right)^2.
    \end{aligned}
    \label{eq:Staggered_hamiltonian_integrated}
\end{align}
Finally, this expression can again be translated to qubits by using the Jordan-Wigner transformation~\cite{jordan_wigner} (for details, see Appendix~\ref{wilson_operators}). Note that in the staggered case, each site is only populated by a single-component fermionic field. Thus, in contrast to the Wilson case, for $N$ physical sites we obtain a system with $N$ qubits and the dimensionless Hamiltonian reads~\cite{hamer97}
\begin{align}
    \label{Staggered_hamiltonian}
        \begin{aligned}
        W_S &= \frac{x}{2}\sum_{n=0}^{N-2} \left(X_n X_{n+1}+ Y_n Y_{n+1}\right) \\
        &+\frac{m_\text{lat}}{g}\sqrt{x}\sum_{n=0}^{N-1} (-1)^n Z_n
        + \sum_{n=0}^{N-2} \left(l_0 + \sum_{k=0}^{n}Q_k\right)^2 
        \end{aligned}
\end{align}
where $x=1/(ag)^2$ is again the inverse lattice spacing in units of the coupling squared, and $X_n$, $Y_n$, and $Z_n$ are the Pauli matrices acting on spin $n$ as in the Wilson case. The staggered charge operator is given by $Q_n = (Z_n + (-1)^n)/2$, from which we can infer that on even (odd) sites we have a positron (electron) present when the spin is up (down).

The observables for detecting the first-order phase transition can again be straightforwardly expressed in spin language. As opposed to Wilson fermions, in the staggered discretization, the charge configuration on the lattice exhibits a staggering effect, which leads to a non-uniform electric flux on the links, in particular for smaller lattices. Therefore, to suppress the boundary effects as well as the staggering effect while computing the electric field density, two adjacent links in the center of the system are averaged rather than just a single one for the case of Wilson fermions. The corresponding electric field operator including the background field is given by
\begin{align}
    \label{efd_staggered}    
    \begin{split}
        L_S &= l_0 + \frac{1}{2}\left(L_{N/2-2}+L_{N/2-1}\right) \\
        &= l_0 + \frac{1}{2}\left(\sum_{k=0}^{N/2-2} Q_k + \sum_{k=0}^{N/2-1} Q_k\right) \\
        &= l_0 + \frac{1}{4} + \frac{1}{2}\sum_{k=0}^{N/2-2} Z_k + \frac{1}{4}Z_{N/2-1}.
    \end{split}
\end{align}
For the particle number we find
\begin{align}    
    P_S = \frac{N}{2} + \frac{1}{2}\sum_{n=0}^{N-1}(-1)^nZ_n.
    \label{particle_number_staggered_operator}
\end{align}
When there are no particles present, i.e.\ the state of the system is given by spin down on even sites and spin up on odd sites, the sum in the equation above contributes a $-1$ for every site. In contrast, for even sites with spin up and on odd sites with spin down, the sum contributes a $+1$. Hence, the first term in Eq.~\eqref{particle_number_staggered_operator} is added to render $P_S$ positive semidefinite. 

In summary, both staggered and Wilson fermions provide a viable discretization of the continuum Schwinger Hamiltonian. For a lattice system with $N$ physical sites the Wilson discretization results in a Hamiltonian on $2N$ qubits, whereas the staggered approach only requires $N$ qubits. While both formulations reproduce the correct continuum limit, as we demonstrate explicitly in Sec.~\ref{results}, it is a priori not clear which of the two discretizations converges faster. Although the staggered approach requires less qubits than the Wilson one, this does not imply it will produce better results given a fixed amount of resources. The latter question is particularly relevant for quantum computing, as current and near-term devices only offer a limited number of qubits that still suffer from a considerable level of noise. Here, we aim at testing both approaches in a realistic scenario to benchmark their performance. For the rest of the paper we will focus on the sector of vanishing total charge, $\sum_n Q_n=0$, for both approaches.

\section{Methods\label{methods}}

In order to assess both fermion discretizations, we study their performance with a VQE as well as their convergence towards the continuum limit. Here, we introduce the VQE setup we consider, including a description of the parametric ansatz circuits and the optimization procedure for the parameters we utilize. Moreover, we discuss the error mitigation techniques used for the inference runs on quantum hardware. Finally, we briefly describe the MPS techniques we use to explore the behavior of both discretizations towards the continuum limit.

\subsection{Parametric ansatz circuits for VQE\label{sec_ansatz}}

\begin{figure}[htp!]
    \centering
    \includegraphics[width=0.98\columnwidth]{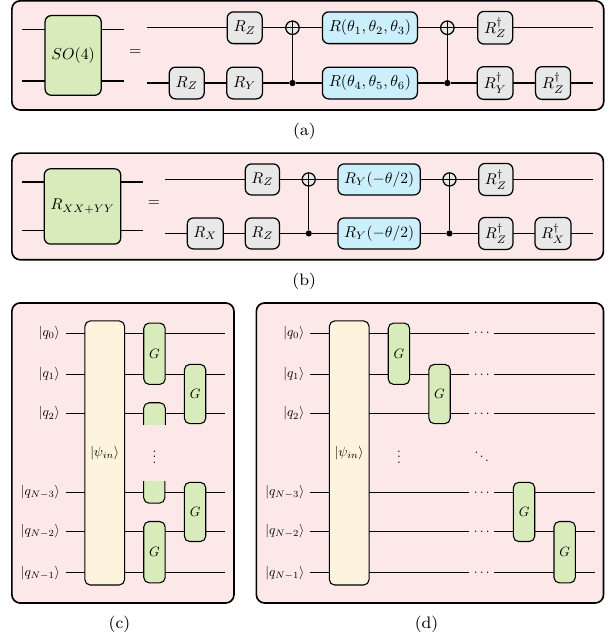}
    \caption{Decomposition of a generic SO(4) gate depending on the six parameters $\theta_1,\dots,\theta_6$ (a) the $R_{XX+YY}(\theta) = R_{Z_{0}}\exp(-i\theta(XX+YY)/2)R_{Z_{0}}^{\dagger}$ (b), into CNOT and Pauli rotation gates. The $R_{Z_{0}}$ rotations in the definition of $R_{XX+YY}(\theta)$ restrict the state to the real subspace. Boxes acting on a single qubit correspond to Pauli rotation gates, $R_P(\alpha) = \exp{-i\alpha P/2}$ with $P\in\{X,Y,Z\}$. Single-qubit gates where the argument is omitted refer to rotations around an angle $\pi/2$, $R_P(\pi/2)$. The light blue boxes represent the parameterized gates which are $R(\alpha, \beta, \gamma) = R_X(\gamma)R_Z(\beta)R_X(\alpha)$ in (a) and $R_Y(\theta)$ in (b). Panel (c) and (d) illustrate one layer of the brick and ladder ansatz, respectively, both following a non-parametric part for preparing the initial state $\ket{\psi_{in}}$ (yellow box). The first layer in the brick ansatz has a CNOT-depth of $4$  whereas in ladder it is $2n - 2$, where $n$ is the number of qubits, and in both cases it increases by $4$ with each layer.}
    \label{ansatz_fig}
\end{figure}

In order to test the performance of the different discretizations for VQE, we focus on two different types of parametric ans\"atze and consider two different types of gates, as shown in Fig.~\ref{ansatz_fig}. We refer to the two ansatz architectures as ``brick'' (c.f.\ Fig.~\ref{ansatz_fig}(c)) and ``ladder'' (c.f.\ Fig.~\ref{ansatz_fig}(d)). The two types of parametric gates we consider are the SO(4) gates and the $R_{XX+YY}$ gates, whose decomposition in standard controlled-NOT (CNOT) and Pauli rotation gates is shown in Figs.~\ref{ansatz_fig}(a) and \ref{ansatz_fig}(b). Here, we have chosen SO(4) instead of SU(4) gates, because the Hamiltonians we study are real, and hence their ground states are real. Thus, we can restrict our ansätze to the real subspace of the Hilbert space. While the SO(4) gates are in principle more expressive than the $R_{XX+YY}$ ones, they do not conserve the total charge. As a result, if we use SO(4) gates, we need to manually enforce vanishing total charge, which we do by adding a penalty term $\lambda\left(\sum_{n=0}^{N-1}Q_n\right)^2$ to the Hamiltonians in Eq.~\eqref{wilson_hamiltonian} and Eq.~\eqref{Staggered_hamiltonian}. The Lagrange multiplier $\lambda$ has to be chosen sufficiently large that one obtains a ground state with vanishing total charge. In contrast, the $R_{XX+YY}$ gate preserves the total charge, but is generally less expressive.
In all our simulations, we choose the initial parameters for the parametric part of this ansatz randomly in the interval $[0, 0.001)$, such that the parametric part of the circuit is close to the identity and we start the VQE with a state close to $\ket{\psi_{in}}$.

For Wilson fermions, three options for the ansatz's non-parametric portion $\ket{\psi_{in}}$ are tested, and the one producing the best fidelity in the VQE will be adopted. As explained in Sec.~\ref{Continuum_Schwinger_Model}, for $m/g\gg1$ and $\theta <\pi$, we expect the continuum Hamiltonian to be dominated by the mass term, and the ground state corresponds to a state with no particles present. The lattice analog of this state for the Wilson case is given by the product state $\left(({\ket{01} - \ket{10}})/\sqrt{2}\right)^{\otimes N}$. As one can easily verify, this state contains no charges and  should provide a good initial state for large masses before the onset of the first-order phase transition. For $\theta > \pi$ the continuum model predicts that a particle antiparticle pair is produced and a negative (positive) charge will form on the left (right) boundary of the system. On the lattice, this would correspond to the state $\ket{11}\left(({\ket{01} - \ket{10}})/\sqrt{2}\right)^{\otimes (N-2)}\ket{00}$ in spin formulation, which approximates the state expected after the phase transition. Finally, we also try a generic state with vanishing total charge given by $\ket{10}^{\otimes N}$. In all cases, a simple linear mapping of the logical qubits $\ket{\sigma_l}$ in the model Hamiltonian to the physical ones of the hardware $\ket{q_l}$ is used.

For the staggered fermions, the state $\ket{10}^{\otimes N/2}$ corresponds to the ground state at large values of $m/g\gg1$ in the sector of vanishing total charge and, thus, is a viable initial state $\ket{\psi_{in}}$ for large masses and $\theta < \pi$. For $\theta>\pi$ we expect similar to the Wilson case a pair of charges to form with the negative (positive) charge located on the left (right) boundary of the system. Due to the staggered formulation, negative (positive) charges can only reside on odd (even) sites, thus the pair will be located at sites $1$ and $N-2$, resulting in the state $\ket{11}\ket{10}^{\otimes (N-2)/2}\ket{00}$. This state can be easily created from $\ket{10}^{\otimes N/2}$ by acting with an SO(4) or an $R_{XX+YY}$ gate between qubits $1$ and $N-2$. For the ansatz architectures shown in Fig.~\ref{ansatz_fig}, there is no two-qubit gate directly acting on qubits $1$ and $N-2$. In order to provide such a connection between these qubits, we deviate from the usual linear mapping of the logical qubits in the model Hamiltonian to the physical ones of the hardware. In particular, we choose the mapping $\ket{q_0}\otimes \ket{q_1} \otimes \dots  \otimes \ket{q_{N-1}}\equiv \ket{\sigma_{(N-2)/2 +1}}\otimes \dots\otimes\ket{\sigma_{N-3}} \otimes\ket{\sigma_{N-1}}\otimes\ket{\sigma_{N-2}} \otimes \ket{\sigma_{1}}\otimes\ket{\sigma_{0}}\otimes\ket{\sigma_{2}}\otimes \dots \otimes\ket{\sigma_{(N-2)/2}}$, where $\ket{q_k}$, $q_k\in\{0,1\}$ are the basis states of the qubits $k$, and $\ket{\sigma_{l}}$, $\sigma_l\in\{0,1\}$ represent the basis states of the logical qubit $l$ in the Hamiltonian. Using the ans\"atze shown in Fig.~\ref{ansatz_fig} on the physical qubits, this provides a direct two-qubit gate between the logical qubits $1$ and $N-2$. Thus, for the staggered case with large masses, we use the initial state $\ket{10}^{\otimes N/2}$ for the logical qubits and then proceed with the mapping discussed above. In the opposite regime, $m/g\ll 1$, in which we expect the ground state to (partially) screen the electric field, we do not foresee that a deviation from the linear mapping is required. Thus in this parameter range, we choose $\ket{10}^{\otimes N/2}$ as an initial state for the logical qubits and simply use the linear mapping as in the Wilson case.

In order to benchmark the ans\"atze for the different discretizations, we first simulate the VQE classically. We employ the L-BFGS-B optimizer~\cite{lbfgs} with two distinct heuristic warm-start stages. In the first phase, we group all variational parameters in a layer and collectively optimize them using a VQE with 2000 iterations. This stage is based on the idea that for large enough system sizes the model should exhibit translation invariance. While translation invariance is broken in the presence of open boundary conditions, this will still provide a reasonable approximation. In the second phase, each parameter can be independently varied in a VQE. At this stage, we allow for an extremely large number of 100,000 iterations to assess the best possible performance of each ansatz. In practice, we typically observe convergence after only a few hundred iterations in most cases. As we increment the layer count from $k$ to $k+1$, the parameters obtained from the prior warm-start phase for $k$ layers are used again in the first $k$ layers to being the warm start for $k+1$ layers. 

To demonstrate that our VQE ansatz is also suited for current quantum devices, we take in a second step the parameters obtained from the classical simulations and perform inference runs on a quantum device. In particular, we prepare the corresponding states on IBM's superconducting quantum hardware and measure the electric field density as well as the particle number. We explore a wide range of parameter regimes ranging from weak coupling, $m/g\gg 1$, in which we expect the first-order phase transition to happen up to $m/g\ll 1$, in which limit we do not expect any transition. In order to obtain the best possible hardware results, we utilize several error mitigation techniques, which we outline in the next section.

\subsection{Error mitigation techniques\label{error_mitigation_techniques_subsection}}

In order to mitigate various errors occurring on the quantum devices we use, we employ a composite mitigation strategy combining several techniques. The backbone of our composite approach is zero-noise extrapolation (ZNE)~\cite{ZNE, majumdar2023best}. We use digital circuit folding, i.e.\ inserting pairs of unitaries $U^\dagger U$ in the circuit, which would result in an identity operation on an ideal quantum computer. For real quantum devices with noise, this effectively allows for running the same circuit at different (and larger) noise levels. Subsequently, the results can be extrapolated to zero noise with an appropriately chosen fitting function~\cite{ZNE}. While ZNE allows for addressing incoherent errors, it cannot mitigate coherent errors.

To counter coherent errors, we employ Pauli twirling to two-qubit CNOT gates and to measurements, in order to transform their coherent quantum error channels into incoherent ones~\cite{PhysRevA.94.052325, pauli_twirling_for_coherent_noise, trex}. A two-qubit gate can be twirled by adding random single-qubit Pauli gates before and after it in a way that the action of the resulting operation is logically unchanged compared to the original one. Sequential single-qubit gates resulting from twirling overlapping CNOT gates can be merged to minimize the gate count in the circuit.  Averaging the results over many random instances of twirled circuits turns coherent errors into incoherent ones. In our case, we use 10 twirls and average over their result for each noise factor in zero-noise extrapolation. Measurements can be twirled in a similar manner, except that a random single-qubit Pauli is inserted only before the measurement. The logical change of basis is then corrected in post-processing. Different instances of twirled circuits can be generated by sampling new single-qubit Pauli gates. 

Lastly, dynamical decoupling is harnessed to tackle decoherence. During extended idling periods comparable to the decoherence time, crosstalk between neighboring qubits can cause decoherence. By leveraging externally controllable interactions, such as spin flip cycles, during these idle intervals, we can suppress this form of decoherence. Here, we use the $XX$ sequence, or two X pulses placed uniformly in all idle times~\cite{dynamical_decoupling}.

\subsection{Extrapolation to the continuum limit with Matrix Product States\label{methods_section_matrix_product_states}}

In order to estimate the system sizes required to obtain a faithful continuum limit and to examine the behavior of both fermion discretizations towards that limit, we need to reach system sizes that cannot directly be simulated on a classical computer. To go beyond the limit of a few tens of qubits, we use MPS. MPS are an entanglement-based ansatz for quantum states~\cite{Orus2014a, Bridgeman2017, Schollwoeck2011, mps_review_2} that can efficiently represent ground states of local gapped Hamiltonians~\cite{Hastings_2007}. For a system of $N$ sites with open boundary conditions, the MPS ansatz reads
\begin{align}    
        \ket{\psi} =  \sum_{\sigma_{0},..,\sigma_{N-1}} A^{\sigma_{0}}_0A^{\sigma_{1}}_1\dots A^{\sigma_{N-1}}_{N-1} \ket{\sigma_{0}} \otimes \ket{\sigma_{1}} ... \otimes \ket{\sigma_{N-1}},
        \label{eq:mps_obc}
\end{align}
where $\{\ket{\sigma_{k}}\}_{k=1}^d$ is a local basis for each site, and the $A^{\sigma_{k}}_k$ are complex $D\times D$ matrices for $0<k<N-1$. The first (last) tensor $A^{\sigma_{0}}_0$ ($A^{\sigma_{N-1}}_{N-1}$) is a $D$-dimensional row (column) vector. For a fixed combination of physical indices, the coefficient of the wave function is parameterized as a product of matrices, hence the name MPS. The size of the matrices $D$ determines the amount of variational parameters that are present in the ansatz and limits the maximum entanglement in the state~\cite{Orus2014a, Bridgeman2017, Schollwoeck2011, mps_review_2}. 

The MPS approximation for the ground state of a given Hamiltonian can be found with standard variational algorithms. The tensors are updated iteratively one by one, while keeping all others fixed. The optimal tensor at each step is obtained by finding the ground state of an effective Hamiltonian that describes the interaction of the tensor with its environment. Repeating the procedure starting from one of the boundaries and sweeping back and forth until the relative change of the energy is below a certain tolerance, the resulting MPS represents an approximation for the ground state of the Hamiltonian. 

MPS algorithms routinely deal with several hundreds of spins and allow us to check the system sizes that are required for obtaining a reliable continuum extrapolation for both fermion discretizations. In particular, we use systems with $N \in \{70, 80, 90, 100\}$ physical sites and focus on the continuum limit at a dimensionless lattice volume $N/\sqrt{x} = 30$. These system sizes seem to be within reach on near-term quantum devices.

\section{Results\label{results}}

Having summarized the methods in Sec.~\ref{methods} that we have used to explore the lattice Schwinger model, we examine the performance of the ansätze for the VQE using classical simulations. Furthermore, we demonstrate the first-order phase transition with quantum hardware using the ground states obtained from the classical simulation of the VQE. Finally, we perform the extrapolation to the continuum with data obtained from an MPS calculation and estimate the resource required for a reliable continuum limit.

\subsection{VQE results\label{vqe_results_section}}

In order to explore the capacity of our ansatz of faithfully representing the ground states, the VQE is simulated classically in the absence of any noise following the procedure outlined in the previous section. We focus on two regimes, on the one hand a large value $m_\text{lat}/g=10$, for which we expect the first-order quantum phase transition to occur, and on the other hand $m_\text{lat}/g=0$, which should be considerably below the critical mass and, thus, no transition occurs as we change $l_0$. For the rest of this section, we consider a fixed lattice volume of $N/\sqrt{x} = 30$.

Focusing on the case of case of Wilson fermions first, we generally observe a similar performance of both the brick and the ladder ansatz. The choice of SO(4) gates mostly leads to a slightly better fidelity than the $R_{XX+YY}$ gates, independently of the mass. Since the brick ansatz has a smaller CNOT-depth than the ladder one for the same number of layers, we deem it more suitable for near-term quantum hardware, and we focus on the brick ansatz with SO(4) gates for the Wilson case. Using the two product states described in Sec~\ref{sec_ansatz} as initial states before and after the transition for the Wilson discretization with $m_{\text{lat}}/g = 10$, the VQE was able to achieve fidelities above 0.99 with a single layer of the brick ansatz with SO(4) gates for all system sizes we study. Figure~\ref{efd_vs_l_0_real_hardware_comparison_mass_10} shows the results from the VQE for electric field density, $\langle L_{W}\rangle$, (first row of Fig.~\ref{efd_vs_l_0_real_hardware_comparison_mass_10}) and the particle number, $\langle P_W\rangle$, (third row of Fig.~\ref{efd_vs_l_0_real_hardware_comparison_mass_10}) as a function of $l_0$ for system sizes $N=3$, $4$, $5$, $6$, which correspond to 6, 8, 10 and 12 qubits for Wilson fermions. In particular, the results for the VQE show the distinct discontinuity in the electric field density and the particle number, indicating the remnant of the first-order phase transition. The location of the transition is shifted to values of $l_0$ much larger than the prediction for the continuum model, $l_0=1/2$. As we increase $N$, we observe that the transition point moves closer towards smaller values of $l_0$. This behavior can be explained with the effects of the finite volume and lattice spacing on the transition point, which we discuss in detail in Appendix~\ref{prediction_for_transition}. More specifically, since we increase $N$ and keep the volume constant, we effectively reduce the lattice spacing, hence going closer to the continuum result. Comparing the results from the VQE to the ones obtained from exact diagonalization, we observe excellent agreement, reflecting the high fidelities that were reached. This can be accounted for by the fact that for $m_{\text{lat}}/g = 10$ the true ground states before and after the transition are close to the product states we use as initial states. As a result, one layer of the ansatz is sufficient to generate essentially perfect overlap.

Turning to a smaller lattice mass of $m_{\text{lat}}/g = 0$, the ground state is more complicated, and in general our VQE needs more layers to achieve equally good fidelities as previously. For the most part, two layers of the brick ansatz with SO(4) gates gave the best performance resulting in fidelities above 0.99. In a few occasions, the $R_{XX+YY}$ gates showed a slightly better performance, and we observed a few instances where a single ansatz layer was already sufficient to reach fidelities above 0.99. Figure~\ref{efd_vs_l_0_real_hardware_comparison_mass_0} summarizes the results for the electric field density and the particle number for $m_{\text{lat}}/g = 0$. Compared to the larger mass, we now see that both quantities exhibit a smooth behavior, as one would expect for masses below the critical one. It can be seen that for a few values of $l_0$, the VQE did not fully converge to the exact diagonalization, especially for larger values of $N$ close to $l_0=1/2$ (see Figs.~\ref{efd_vs_l_0_real_hardware_comparison_mass_10}(b), \ref{efd_vs_l_0_real_hardware_comparison_mass_10}(c), \ref{efd_vs_l_0_real_hardware_comparison_mass_10}(d) and \ref{efd_vs_l_0_real_hardware_comparison_mass_10}(j)). In theses cases the best fidelity obtained is around 0.85. Looking at the electric field density and the particle number for larger $N$, we observe a rather larger slope around $l_0=1/2$. This indicates that despite choosing a small lattice mass, we are not too far away from the first-order phase transition, possibly due to finite-size effects and the additive mass renormalization. Hence, at $l_0=1/2$ there are presumably two states with similar energy making it harder for the VQE to converge. Nevertheless, our classically simulated VQE results are in general in good agreement with the data from exact diagonalization for a wide range of parameters.

Focusing on the case of staggered fermions and using the initial states and qubit mappings discussed in Sec.~\ref{sec_ansatz}, the ans\"atze with the $R_{XX+YY}$ gates generally returned higher fidelities compared to the ones with the $SO(4)$ gates. In the latter case, we observed the tendency that the final parameters obtained by the optimization routine at the end of the VQE were all close to $0$, resulting in a state formed only from the initial non-parametric part (which prepares the $\ket{10}^{\otimes N/2}$ state). In contrast, with $R_{XX+YY}$ gates both of the architectures performed comparably well and were able to achieve fidelities above $0.99$. Since the brick architecture has smaller CNOT-depth, we again focus on this one for the staggered discretization. 
Considering $m_{\text{lat}}/g = 10$ first, the staggered discretization behaves qualitatively similar to the Wilson fermions, as Fig.~\ref{efd_vs_l_0_real_hardware_comparison_mass_10} reveals. We again observe the characteristic discontinuities in the electric field density (second row of Fig.~\ref{efd_vs_l_0_real_hardware_comparison_mass_10}) and the particle number (fourth row of Fig.~\ref{efd_vs_l_0_real_hardware_comparison_mass_10}), indicating the first-order quantum phase transition. Comparing to the Wilson fermions, the transition for the staggered discretization occurs at larger $l_0$, despite the fact that for both discretizations we match the qubit number, which effectively means that for the staggered fermions we have a larger number of physical sites and, since the volume is fixed to $N/\sqrt{x}=30$, a finer lattice spacing than for the Wilson fermions. This effect might be caused by a larger additive mass renormalization compared to the Wilson case. Also for the staggered discretization, we see the same trend that with increasing system size it is shifted to smaller values of the $\theta$-angle (for details, see Appendix~\ref{prediction_for_transition}). In particular, the data from exact diagonalization is in excellent agreement with the one from simulated VQE with one ansatz layer, thus showing that our ansatz circuit together with the qubit mapping described in Sec.~\ref{methods} is appropriate for capturing the physics of the model in the regime of large masses.

Going to the opposite regime of small lattice mass, $m_{\text{lat}}/g = 0$, we again have to increase the number of layers to two for the staggered case to have high fidelities with the exact results, due to the more complicated nature of the ground state in this regime. The results for two layers of the brick ansatz are shown in Fig.~\ref{efd_vs_l_0_real_hardware_comparison_mass_0}. Similar to the Wilson fermions, also for the staggered discretization, the electric field density (particle number) now shows a smooth decrease (increase), indicating that the absence of the first-order phase transition for the chosen lattice mass. Interestingly, for the staggered discretization we essentially obtain perfect agreement with the data from exact diagonalization, and there the results from classically simulated VQE do not show any deviations for the entire parameter range we study.

In summary, our results for the Wilson case suggest that the brick ansatz with SO(4) gates with an appropriate initial state is generally favorable, and that the resources required for approximating the ground state to a good level of accuracy do not show a strong dependence on the system size in the range we study. For the staggered case, our data imply the brick ansatz with $R_{XX+YY}$ gates provides good performance using the same initial state for the entire parameter regime we explore. Moreover, also for the staggered case there is no strong dependence on the system size in the range we study.

\subsection{Inference runs on quantum hardware}
\label{quantum_hardware_results}

\begin{figure*}
    \centering
    \includegraphics[width = \textwidth]{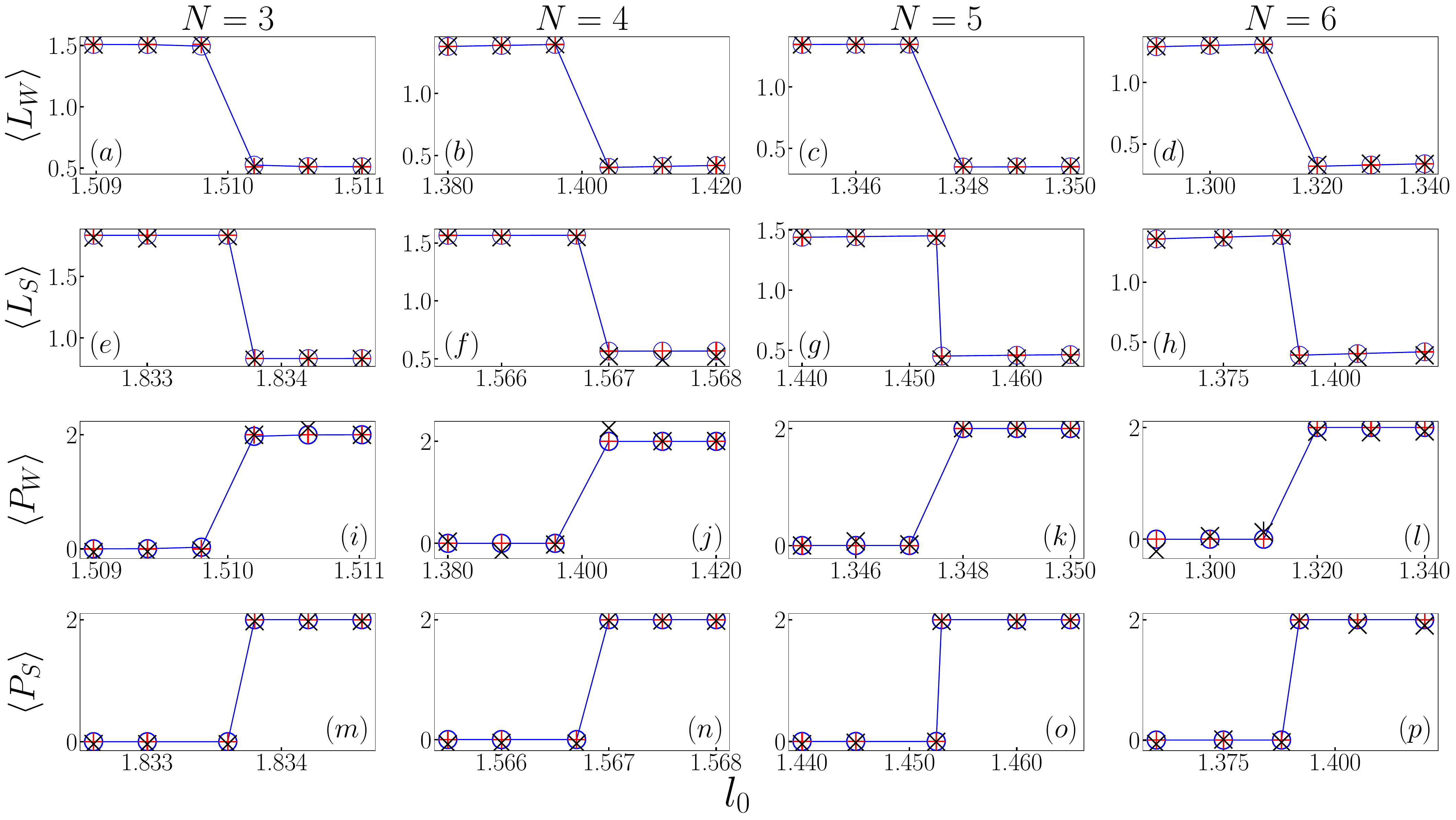}
    \caption{Electric field density $\langle L_{W,S} \rangle$ and particle number $\langle P_{W,S} \rangle$ against $l_0$ with data from quantum hardware (black crosses), as compared to the noiseless expectation values (red pluses) and exact diagonalization (blue circles) for Wilson and staggered fermions respectively. The staggered fermions in this case have the same number of qubits as the Wilson fermions. Hence, while the title of each column specifies the $N$ for Wilson fermions, for staggered it is taken to be double that value. The lattice mass for these data is set to $m_{\text{lat}}/g = 10$, so that we are above the second order phase transition of Fig.~\ref{phase_diagram}, without having to account for the mass shift. Therefore, we can observe the first-order phase transition. Note that the error bars, which are discussed in Appendix~\ref{zne_plots_appendix}, are much smaller than the y-scale and thus, are not visible.}
    \label{efd_vs_l_0_real_hardware_comparison_mass_10}
\end{figure*}

\begin{figure*}
    \centering
    \includegraphics[width = \textwidth]{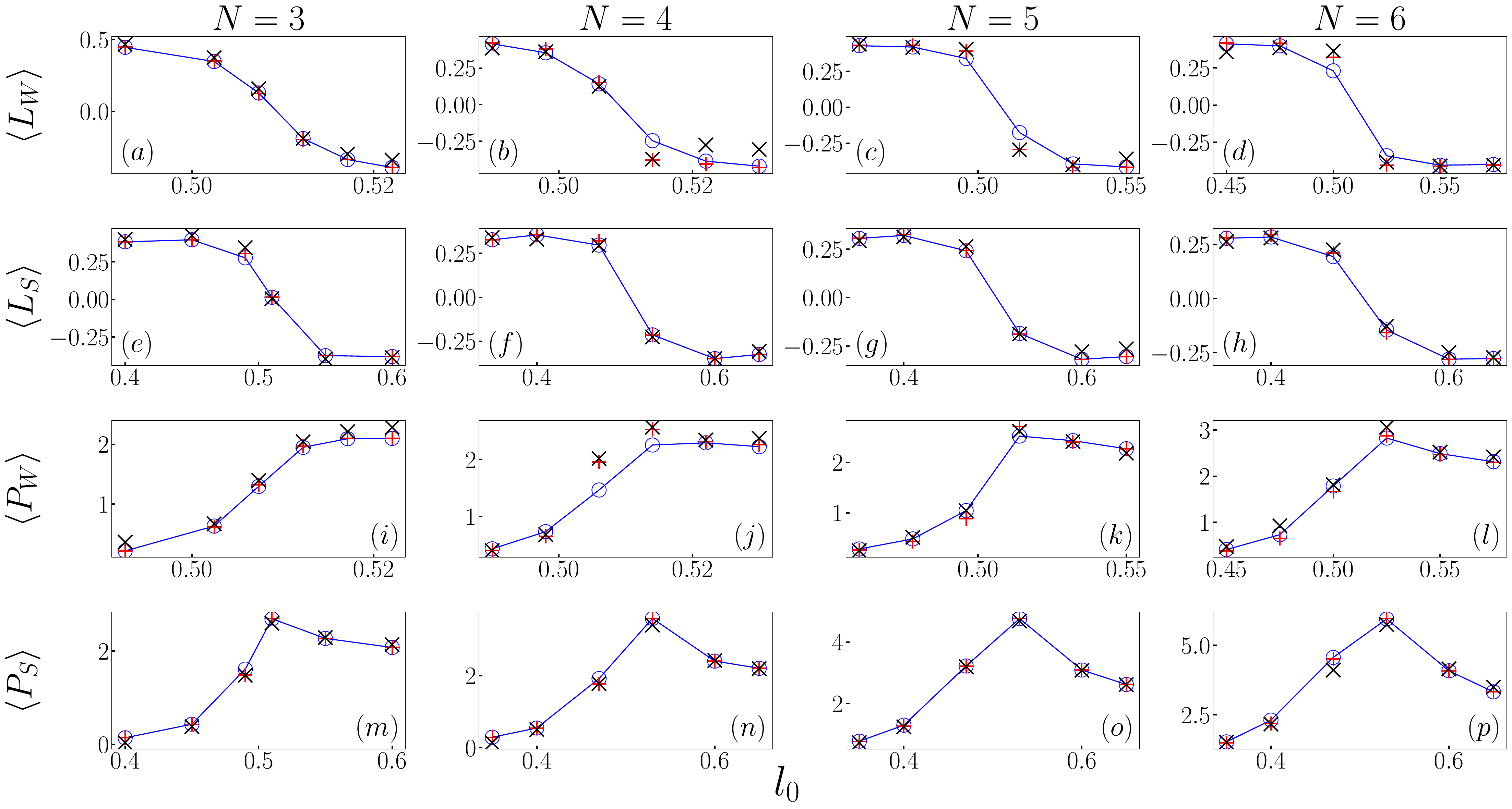}
    \caption{The description for this figure follows Fig.~\ref{efd_vs_l_0_real_hardware_comparison_mass_10}, however the lattice mass here is set to $m_{\text{lat}}/g = 0$, so that we are below the second order phase transition and thus observe no first-order phase transition as expected. Note that the error bars are much smaller than the y-scale and thus, are not visible. The blue circles represent the exact diagonalization, red pluses the noiseless simulations and black crosses the quantum hardware results.}
    \label{efd_vs_l_0_real_hardware_comparison_mass_0}
\end{figure*}

The previous section has shown that our VQE ansatz is able to capture the relevant ground states efficiently with a small number layers. To demonstrate the approach is feasible on current and near-term quantum hardware, we perform inference runs on quantum hardware. To this end, we prepare the ansatz circuit for the parameters obtained at the end of the classical simulation of the VQE on a quantum device and measure the electric field density as well as the particle number. We use IBM's quantum devices \textit{ibm\_hanoi}, \textit{ibm\_cusco} and \textit{ibm\_nazca} for our inference runs, where we perform $10^4$ measurements in the computational basis. The results for both fermion discretizations after applying the error mitigation techniques discussed in Sec.~\ref{error_mitigation_techniques_subsection} are shown in Figs.~\ref{efd_vs_l_0_real_hardware_comparison_mass_10} ($m_{\text{lat}}/g = 10$) and in \ref{efd_vs_l_0_real_hardware_comparison_mass_0} ($m_{\text{lat}}/g = 0$), alongside with the data from the simulated ideal VQE and the results from the exact diagonalization. 

In general, the results from the quantum hardware agree well with the noise-free simulation of the VQE, except at a few points. This discrepancy may be a result of several factors. We have observed for example that measurements performed right after calibration of the quantum hardware were closer to the simulated result, as compared to measurements done when the last calibration took place several hours before. Furthermore, the case $m_{\text{lat}}/g = 0$ has a few points which do not match well with the corresponding noise-free ones, compared to the larger mass of $m_{\text{lat}}/g = 10$. This can be related to the fact that the circuits for $m_{\text{lat}}/g = 0$ were one layer deeper, which generally leads to larger effects of noise on the results.

\begin{figure*}
    \centering
    \includegraphics[width = \textwidth]{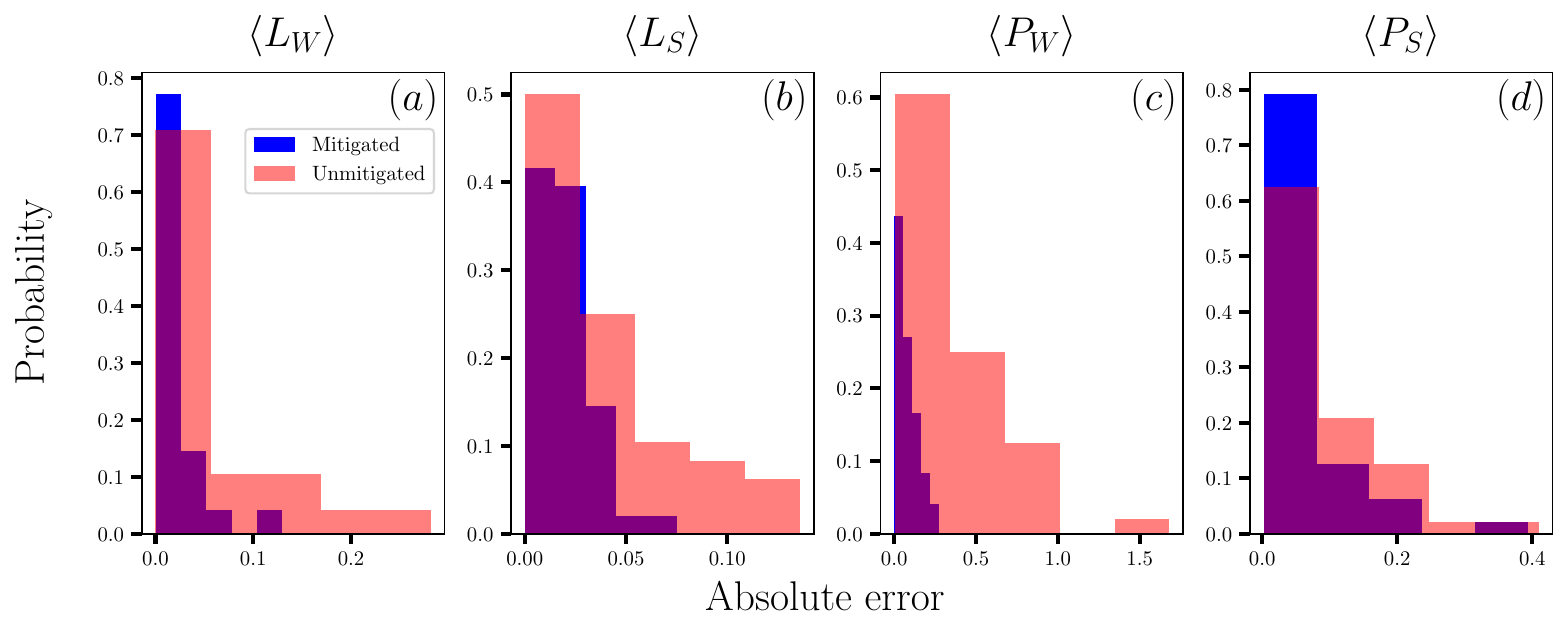}
    \caption{Histograms for the absolute errors of the mitigated and unmitigated data for the electric field density for the Wilson (a) and the staggered discretization (b) as well as the particle number for the Wilson fermions (c) and the staggered formulation (d). Blue bars represent the absolute error between the hardware data after ZNE and the simulated VQE, while red bars represent the absolute error between unmitigated hardware data and the simulated VQE. Each subplot includes all $N, m_{\text{lat}}/g, l_0$.}
    \label{histograms_real_hardware_comparison}
\end{figure*}

To assess the performance of the error mitigation procedure, we show in  Fig.~\ref{histograms_real_hardware_comparison} histograms for the absolute error of the unmitigated and mitigated results for the particle number and the electric field density. The absolute error (shown on the $x$-axis) corresponds to the absolute value of the difference between the ones obtained from the simulated VQE, and the (un)mitigated values from the quantum device. For the electric field density in the Wilson case and the particle number in the staggered formulation, the histograms in Figs.~\ref{histograms_real_hardware_comparison}(a) and ~\ref{histograms_real_hardware_comparison} (d) show a clear improvement. The mitigated values (blue bars) are concentrated around smaller errors and have a higher probability of coinciding with the correct result than the unmitigated ones (red translucent bars). For the electric field density in the staggered case and the particle number in the Wilson discretization, we observe a different picture. Figures~\ref{histograms_real_hardware_comparison}(b) and ~\ref{histograms_real_hardware_comparison}(c) show that the probability of measuring the exact result has decreased compared to the unmitigated results. Nevertheless, the overall distribution after mitigation shows a smaller width and is concentrated at smaller values of the absolute error than the original data. This effect can be explained by the fact that the ZNE is in general not guaranteed to improve the results, and occasionally it can make the final result slightly worse (examples of ZNE extrapolations are shown in detail in Appendix~\ref{zne_plots_appendix}). In general, we observe a positive effect of the error mitigation, as the expected error is reduced in all cases.

Having demonstrated, that our ansatz can be reliably executed on current and near-term quantum hardware for system sizes corresponding 6-12 qubits, we examine if a near-term device with $\mathcal{O}(100)$ qubits would allow for reliably studying the continuum limit of the model.

\subsection{Continuum extrapolations with MPS\label{results_with_tensor_networks}}

To investigate the system sizes required for a reliable extrapolation to the continuum limit as well as to compare the performance of both fermion discretizations towards that limit, we use MPS simulations (for details, c.f.\ Appendix~\ref{numerical_tn_data}). We use again a fixed lattice volume of $N/\sqrt{x} = 30$, where we restrict $N \in \{70,80,90,100\}$. This allows us to study the fixed volume continuum limit, $ag\to 0$, with resources that should be accessible on current and near-term quantum hardware. In order to be able to follow a line of constant renormalized mass, towards the continuum limit, we determine the mass shift for each value of $N$ and $l_0$ following the ideas in Ref.~\cite{angelides}. Here we focus on a small value, $m_r/g=0.01$, as for large masses the ground state of the model is closer to a product state and small masses are in general more challenging. In particular, we will focus on the electric field density, as for this observable our continuum extrapolations can be compared to the prediction from mass perturbation theory~\cite{adam}, which reads
\begin{equation}
    \label{adam_efd}
    \frac{\mathcal{F}}{g} = \frac{e^\gamma}{\sqrt{\pi}}\Big(\frac{m}{g}\Big)\sin{\theta} - 8.9139\text{ }\frac{e^{2\gamma}}{4\pi}\Big(\frac{m}{g}\Big)^2\sin{(2\theta)},
\end{equation}
where $\gamma = 0.5772156649$ is the Euler-Mascheroni constant.

\begin{figure}
    \centering
    \includegraphics[width=\linewidth]{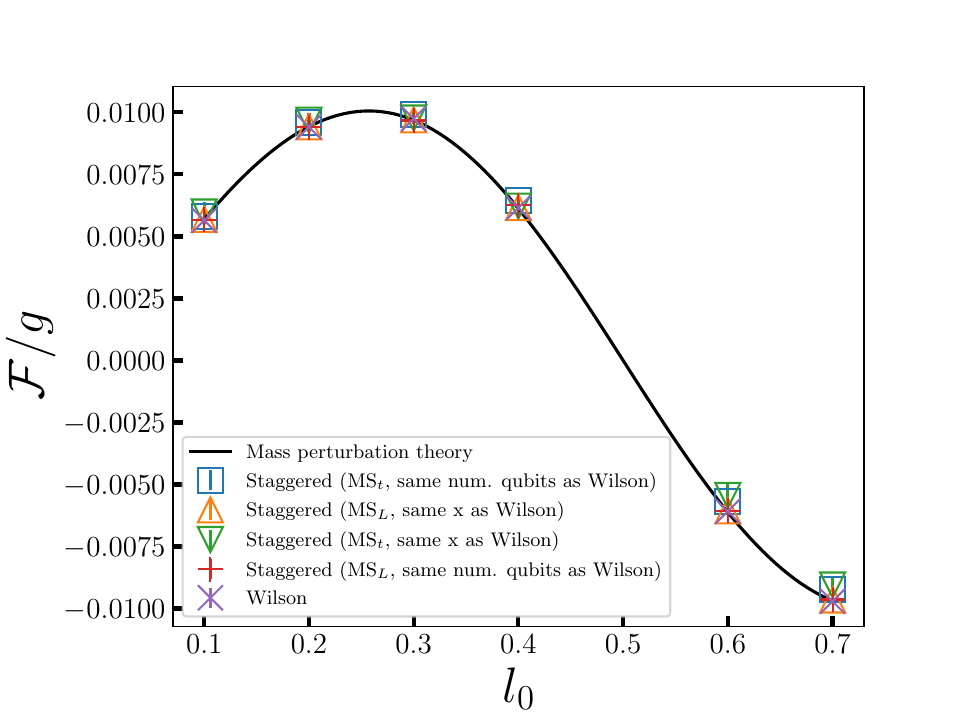}
    \caption{Electric field density $\mathcal{F}/{g}$ against $l_0$ for Wilson fermions and staggered fermions at $m_\text{r}/g = 0.01$. The following MSE values are the mean squared error of each of the data as compared to the continuum mass perturbation theory prediction for the electric field density $\mathcal{F}/g$ given in Eq.~\eqref{adam_efd}. For staggered fermions with $\text{MS}_t$ and same $x$ (number of qubits as Wilson), the MSE is $1.041\times10^{-6}$ ($6.16\times10^{-7}$), and respectively for $\text{MS}_{L}$ $6.143\times10^{-9}$ ($2.617\times10^{-8}$). For Wilson fermions the MSE is $7.926\times10^{-9}$. The error bars emanate from the errors in the variational algorithm to compute the relevant ground states, the extrapolation in bond dimension and in lattice spacing.}
    \label{compare_efd_vs_l_0}
\end{figure}

Figure~\ref{compare_efd_vs_l_0} shows the continuum result for  the electric field density obtained for each discretization in comparison with prediction from mass perturbation theory. To compare the performance between the staggered and the Wilson discretization, there are two options. On the one hand, one can choose the same value of $x = 1/(ag)^2$ for both, which results in a staggered formulation that has half of the number of qubits than the Wilson one. On the other hand, one can choose the same number of qubits for both discretizations, which leads to double the number of physical sites for the staggered formulation, which, given the fixed lattice volumes we work with, simultaneously means a finer lattice spacing than in the Wilson case. For each case, in the staggered formulation we additionally compare the results obtained with using the theoretically predicted values of the mass shift in Eq.~\eqref{eq:ms_staggered}, and the ones from measuring the mass shift $\text{MS}_L$ following Ref.~\cite{angelides}. Hence, there are in total four combinations of the two above considerations, which are all shown in Fig.~\ref{compare_efd_vs_l_0}.

In general, we observe the most precise results from both Wilson and staggered fermion when using $\text{MS}_L$ and the same $x$ as Wilson. When the staggered utilized $\text{MS}_L$ but retained an equivalent qubit count to Wilson, the performance dipped, primarily since resources (like bond dimension $D$) were consistent with the same $x$ scenario, even if the former had an expanded $N$. Further, the staggered fermions using the $\text{MS}_t$ were not able to match the best performance, since the $\text{MS}_t$ does not take into account the $l_0$ dependence to the MS; this effect is exemplified in the points with higher $l_0$ which are further away from perturbation theory in Fig.~\ref{compare_efd_vs_l_0}. Appendix~\ref{numerical_tn_data} includes Table~\ref{compare_efd_vs_l_0_table} which quantifies this deviation by showing the absolute distance between the points in Fig.~\ref{compare_efd_vs_l_0} and mass perturbation theory. An $l_0$ dependence in the massless Schwinger model with staggered fermions was also reported in Fig.~7 of~\cite{Azcoiti}. We emphasize that the $\text{MS}_t$ is derived with periodic boundary conditions while we only consider OBC. 

Overall, we conclude that, even with these modest resources, the extracted data closely aligns with the continuum theory, both qualitatively and quantitatively. However, disregarding the mass shift significantly skews the extrapolated data in comparison to the theory (for details, see Appendix~\ref{numerical_tn_data}). Hence, if this procedure is to be approached with digital quantum computing, the mass shift will play an important role in giving reliable continuum limit results with relatively small system sizes up to $100-200$ qubits.

\section{Conclusion\label{sec:conclusion}}

In this work, we developed VQE ansätze for the lattice Schwinger model suitable for studying its phase structure on current and near-term quantum devices in the presence of a topological $\theta$-term, a regime where conventional Monte Carlo methods suffer from the sign problem. Using two types of fermion discretizations, Wilson and staggered fermions, we demonstrated that for both approaches it is possible to find shallow parametric ansatz circuits that allow for obtaining high fidelities with the exact solution for a large range of system sizes, masses and $\theta$-values. In particular, we demonstrated that we can observe the remnants of the expected phase structure of the first-order phase transition at $\theta=\pi$ occurring in the regime of large masses. Moreover, preparing the resulting wave function for up to $12$ qubits on IBM's quantum hardware, we showed that the characteristic observables revealing the phase structure, the electric field density and the particle number, can be measured precisely on quantum hardware when using state-of-the-art error mitigation techniques, notably zero noise extrapolation, Pauli twirling and dynamical decoupling. In general, our simulations on classical and quantum hardware suggest that both fermion discretizations do not show a large difference in performance for the entire parameter range we study. The ansatz that performed best for the staggered discretization in our studies is slightly simpler than the one for the Wilson fermions. Nevertheless, the results from quantum hardware after error mitigation showed a similar precision for both cases.

Furthermore, we used MPS simulations to estimate the resources for both fermion discretizations required to obtain the continuum limit. Our results show that the continuum values for the electric field density and the particle number can be reliably obtained with $100$ - $200$ qubits, which is within reach with current quantum hardware. In addition, our results demonstrate universality for the two fermion discretizations, as we explicitly observe the same continuum limit for the electric field density and the particle number from both approaches. Comparing both discretizations, for the staggered fermions we observe reduced lattice artifacts towards the continuum limit, thus allowing us to obtain a continuum limit with the same precision at slightly larger lattice spacings compared to Wilson fermions. Moreover, for both cases we observe that taking the additive mass renormalization into account significantly improves the continuum extrapolation.

Given that for the staggered case our MPS simulations showed better convergence towards the continuum limit, and our ansatz circuit for the VQE is slightly simpler, this suggests that the staggered discretization might be easier to address on current and near-term quantum hardware. Combining our results with some of the techniques of Ref.~\cite{2023arXiv230804481F}, it seems possible to reach larger lattice sizes on the order of 100 qubits, and for the future we plan to study the continuum extrapolation directly on quantum hardware. Moreover, it is an interesting question how both fermion discretizations perform in higher dimensions. While for the Schwinger model we integrated the gauge fields out, this is no longer possible for two and more spatial dimensions. This might change the performance of both discretizations and the complexity of a VQE ansatz required to reliably capture the ground state. 

\begin{appendix}

\section{Derivation of the Hamiltonian\label{wilson_operators}}

In this appendix we derive the Hamiltonian for Wilson fermions, in particular, we provide the details of the fermion ordering chosen. Subsequently, we show the Jordan Wigner transformation and the final Hamiltonians in spin formulation used in the simulations for both Wilson and staggered fermions.

We turn first to Wilson fermions starting from the free Dirac Hamiltonian on the lattice which is given by
\begin{equation}
    H = a\sum_n \psi_n^\dagger \gamma^0\left(-i\gamma^1\partial_1 + m_\text{lat}\right)\psi_n,
\end{equation}
with $a$ the lattice spacing, $n$ specifying the lattice site and $m_\text{lat}$ the lattice mass. The derivative $\partial_1\psi_n$ appearing in the above corresponds to the symmetric finite differences $(\psi_{n+1}-\psi_{n-1})/2a$. We add the Wilson term~\cite{angelides, Zache},
\begin{align}
\label{wilson_term}
\begin{split}
    \Delta H_{W} &= -r\frac{a^2}{2}\sum_n\overline{\psi}_n\left(\partial_1\right)^2\psi_n  \\
    &= -r\frac{a^2}{2}\sum_n\overline{\psi}_n\left(\frac{\psi_{n+1}+\psi_{n-1}-2\psi_n}{a^2}\right),
\end{split}
\end{align}
to $H$, where we have used the spatial discrete symmetrized second derivative for $(\partial_1)^2$ and the Wilson parameter $r$. For the $\gamma$ matrices we make the choice $\gamma^0 = X$, $\gamma^1 = iZ$, with $X, Z$, being the standard Pauli matrices. To ensure invariance under local symmetry transformations, we have to introduce the link operators $U_n$, and its canonical conjugate the electric field operator $E_n = gL_n$, with $g$ being the coupling between fermions and bosons. The resulting gauge invariant Hamiltonian reads
\begin{align}
    \label{eq:dirac_hamiltonain_gauged_wilson_fermions}
    \begin{aligned} 
         H_W &= \sum_n\biggr(-\overline{\psi}_n\left(\frac{r+i\gamma^1}{2}\right)U_n\psi_{n+1} \\
         &+ \overline{\psi}_n\left(\frac{-r+i\gamma^1}{2}\right)U^\dagger_n\psi_{n-1} \\
         &+ (am_\text{lat}+r)\overline{\psi}_n\psi_n + \frac{a}{2}\left(E_n + \frac{g\theta}{2\pi}\right)^2 \biggr),
    \end{aligned}
\end{align}
where we explicitly consider the background electric field $g\theta/2\pi$. The gauge invariant states have to fulfill Gauss's law, $L_n - L_{n-1} = Q_n$, with the charge operator given by $Q_n = a\psi_n^\dagger \psi_n - 1$. For open boundary conditions, this expression can be solved iteratively, where we find $L_n = \varepsilon_0 + \sum_{k=0}^{n} Q_k$ with $\varepsilon_0$ the electric field on the left boundary. Similar to the main text, we consider here $\varepsilon_0=0$. Substituting this into the Hamiltonian and applying the unitary transformation from Ref.~\cite{hamer97}, we can fully remove the gauge fields. For numerical calculations, it is convenient to make the fields $\psi$ dimensionless with $\phi_{n,\alpha} = (-1)^n\sqrt{a}\psi_{n,\alpha}$, where the index $\alpha \in \{1,2\}$ specifies the spinor component. We also make the Hamiltonian dimensionless by rescaling it to $\tilde{W} = 2H/ag^2$. Contrary to Refs.~\cite{Zache, angelides}, we additionally apply the transformation $\phi_{n,\alpha} \to (-1)^n(-i)^\alpha\phi_{n,\alpha}$ to the fermionic fields, which results in a completely real Hamiltonian. 

In order translate the fermionic fields to qubits, we choose to map the spinor $\phi_{n,\alpha}$ linearly using the following mapping $\phi_{n,\alpha} \to \chi_{2n-\lfloor\frac{\alpha}{2}\rfloor + 1}$. Now we can apply a Jordan-Wigner transformation~\cite{jordan_wigner}, $\chi_n = \prod_{k<n} (iZ_k)\sigma^-_n$. The charge operator after this transformation becomes $Q_n = (Z_{2n} + Z_{2n+1})/2$, with $n \in \{0, N-1\}$. To enforce vanishing total charge, in case it is needed, we use a penalty term $\Lambda$ to the Hamiltonian, $W_W = \tilde{W} + \Lambda$, where $\Lambda = \lambda\left(\sum_{n=0}^{N-1}Q_n\right)^2$. Provided $\lambda$ is chosen large enough, we can ensure we remain in the sector with vanishing total charge. A few algebraic manipulations on nested sums then results in the final Hamiltonian
\begin{align}
\label{wilson_hamiltonian_final}
    W_W &= x(r-1)\sum_{n=0}^{N-2}\left(\sigma^+_{2n}Z_{2n+1}Z_{2n+2}\sigma^-_{2n+3} + \text{h.c.}\right) \nonumber \\
    &+ \frac{x(r+1)}{2}\sum_{n=0}^{N-2}\left(X_{2n+1}X_{2n+2} + Y_{2n+1}Y_{2n+2}\right) \nonumber \\
    &+ \left(\frac{m_\text{lat}}{g}\sqrt{x} + xr\right)\sum_{n=0}^{N-1}\left(X_{2n}X_{2n+1} + Y_{2n}Y_{2n+1}\right) \nonumber \\
    &+ \frac{1}{2}\sum_{n=0}^{2N-1}\sum_{k=n+1}^{2N-1}\left(N- \left \lceil \frac{k+1}{2} \right \rceil  + \lambda\right)Z_nZ_k \\
    &+ l_0\sum_{n=0}^{2N-3}\left(N- \left \lceil \frac{n+1}{2} \right \rceil \right)Z_n \nonumber \\
    &+ l_0^2(N-1) +\frac{1}{4}N(N-1) + \frac{\lambda N}{2} \nonumber,
\end{align}
where $\lceil \cdot \rceil$ is the ceiling function to the nearest integer greater than the input.

For the staggered case, we directly apply the Jordan-Wigner transformation to Eq.~\eqref{eq:Staggered_hamiltonian_integrated} and set $\phi_{n} =\prod_{k<n} (iZ_k)\sigma^-_n$. Similar to the Wilson case, we add the penalty term to restrict the Hamiltonian to the subspace of vanishing total charge. After the Jordan-Wigner transformation, for staggered fermions, the charge operator becomes $Q_n = (Z_n + (-1)^n)/2$ and the final Hamiltonian including the penalty term is given by
\begin{align}
\label{Staggered_hamiltonian_final}
    \begin{aligned}
        W_S &= \frac{x}{2}\sum_{n=0}^{N-2} \left(X_n X_{n+1}+ Y_n Y_{n+1}\right) \\
        &+\frac{1}{2}\sum_{n=0}^{N-2}\sum_{k=n+1}^{N-1}(N-k-1+\lambda)Z_nZ_k\\
        &+\sum_{n=0}^{N-2}\left(\frac{N}{4}-\frac{1}{2}\left\lceil\frac{n}{2}\right\rceil+l_0(N-n-1)\right)Z_n \\
        &+\frac{m_\text{lat}}{g}\sqrt{x}\sum_{n=0}^{N-1} (-1)^n Z_n\\
        &+l_0^2(N-1) + \frac{1}{2}l_0N + \frac{1}{8}N^2 + \frac{\lambda}{4}N.
    \end{aligned}
\end{align}

\section{Zero noise extrapolations\label{zne_plots_appendix}}

Here we discuss the technical details of the ZNE that we used for our inference runs. In our implementation, global folding was applied with noise factors 1, 3, and 5, with 10,000 shots collected from each circuit. The noise factors are defined as follows: if the circuit is represented by the unitary $U$, then at noise factor $f$, where $f$ is an odd number, a total of $(f-1)/2$ pairs of unitaries $U^\dagger U$ is added to the circuit. For example, noise factor 3, the circuit would be $U(U^\dagger U)$ and at noise factor 5 it would be $U(U^\dagger U)(U^\dagger U)$. 

To get a final zero-noise extrapolated expectation value, we first compute preliminary zero-noise extrapolated values using three approaches: a linear fit to noise factors 1, 3, and 5; a linear fit to noise factors 1 and 3; and a second-order polynomial fit to noise factors 1, 3, and 5. We then calculate the final zero-noise extrapolated value as a weighted average of the preliminary values, weighted according to their variance. Their variance emanates from the fitting procedure performed on points which themselves have a variance. Each point within a fit, i.e. each point at a given noise factor, is the average of the 10 twirls used from the Pauli twirling error mitigation technique described in Sec.~\ref{error_mitigation_techniques_subsection}. Hence, the corresponding error on each point within a fit comes from this average. The error of each one of these 10 twirls before twirl averaging is taken to be the square root of the ratio of the variance of the measurement divided by the number of shots.

\begin{figure}[htp!]
    \centering
    \includegraphics[width=\linewidth]{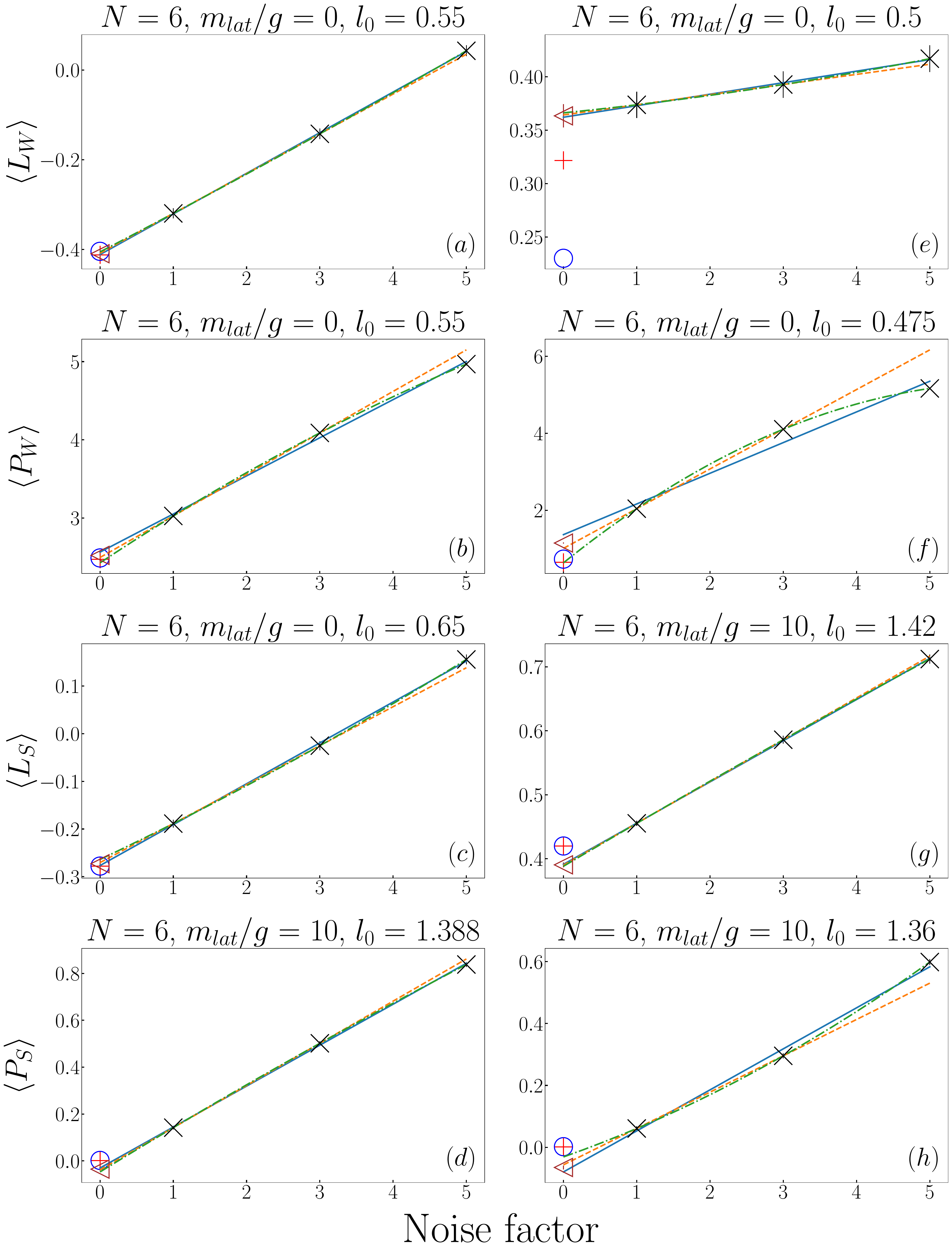}
    \caption{Electric field density $\langle L_{W,S} \rangle$ and particle number $\langle P_{W,S} \rangle$ against noise factor for zero noise extrapolations (ZNE) on the data (black crosses) obtained from quantum hardware for different parameters specified in the titles of each subplot with Wilson and staggered fermions. The red pluses are noiseless values of the observables and the blue circles are the exact diagonalization results. For the fits we fit a first-order polynomial to either all points (blue continuous line) or to points with noise factor 1, 3 (dashed orange line). We also fit a second order polynomial to all points (green dotted dashed line). The brown triangle is the weighted average of the extrapolated points of these 3 fits. The plots on the left $(a-d)$ are showing examples where the ZNE helped take the expectation values measured closer to the noiseless result, while on the right $(e-h)$ we show occasions where the ZNE might not be able to improve the result. However, overall Fig.~\ref{histograms_real_hardware_comparison} indicates that ZNE can significantly improve the results. The calculation of the error bars is described within this Appendix and where they are not visible, they are smaller than the markers and y-scale.}
    \label{specific_zne_fits}
\end{figure}

In Fig.~\ref{specific_zne_fits} we show specific parameter examples of the ZNE fits that contribute to Fig.~\ref{histograms_real_hardware_comparison}. We present representative cases for both situations, where the ZNE was able to improve the expectation value of the observable to be measured and where it was not. Note that Fig.~\ref{histograms_real_hardware_comparison} shows that with respect to all experiments carried out, the ZNE was able to improve the results.

\section{Tensor networks\label{numerical_tn_data}}

In order to find an MPS approximation for the ground states of the two Hamiltonians we consider, we use the two-site variational ground state search algorithm from the ITensors library~\cite{ITensors}. The number of iterations was fixed to 500 and the tolerance $\eta$ for the relative change in energy between two sweeps was set to $10^{-9}$, which in turn gives an approximation for the error on the electric field density given by the value of the electric field density multiplied by $\sqrt{\eta}$~\cite{jutho_thesis}. Additionally, we restrict the MPS to the sector of vanishing total charge by applying the symmetry directly to the ansatz~\cite{global_symmetry_mps}, thus no penalty term is used in the spin Hamiltonians.

\begin{figure}[htp!]
    \centering
    \includegraphics[width=0.9\linewidth]{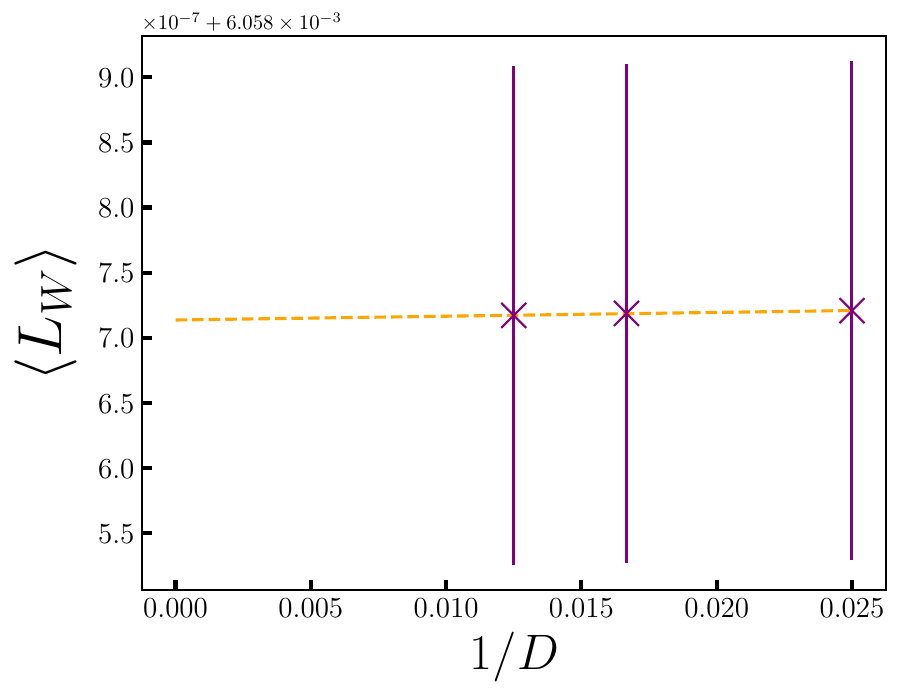}
    \caption{Electric field density $\langle L_W \rangle$ against inverse bond dimension $1/D$ to extrapolate to $D \to \infty$ with $D = 40, 60, 80$. This example is for Wilson fermions at $N = 100$, $l_0 = 0.1$, $m_{\text{lat}}/g = -0.08236266$. Similar behavior was observed for staggered fermions as well. The error bars are emanating from the variational algorithm to compute the relevant ground states.}
    \label{efd_vs_inv_D}
\end{figure}

In order to compare between the Wilson and staggered fermion discretization approaches, we carry out a continuum limit extrapolation. This implies following lines of constant physics in the physical phase space, which requires the calculation of the mass shift for each $N, x, l_0$. With our work confined to a fixed volume of $N/\sqrt{x} = 30$, this necessitates the determination of the mass shift in Eq.~\eqref{mass_shift_equation} for individual values of $N$ and $l_0$ only. In this work, we followed the method based on the electric field density introduced in Ref.~\cite{angelides}. Table~\ref{compare_efd_vs_l_0_table} complements Fig.~\ref{compare_efd_vs_l_0} and quantifies the effect of not accounting for an $l_0$ dependence on the mass shift with staggered fermions. Finally, Fig.~\ref{compare_efd_vs_l_0_Wilson_no_MS} is qualitatively showing the effect of not accounting for the mass shift when using Wilson fermions, which was also observed in~\cite{angelides}. The bond dimension for all MPS data was set to $D = 20, 40, 60, 80$ and we then extrapolate the last 3 points to infinite $D$ in the electric field density against $1/D$ plot as shown in Fig.~\ref{efd_vs_inv_D}. The central value is taken to be the average of the extrapolated value of the linear fit and the electric field density value at the highest $D$ calculated, while the error is approximated to be half the difference between these values. The smaller $D$ solutions were given as initial ansatz to the higher $D$, while $D = 20$ was initialized with a random MPS.

\begin{figure}[htp!]
    \centering
    \includegraphics[width=\linewidth]{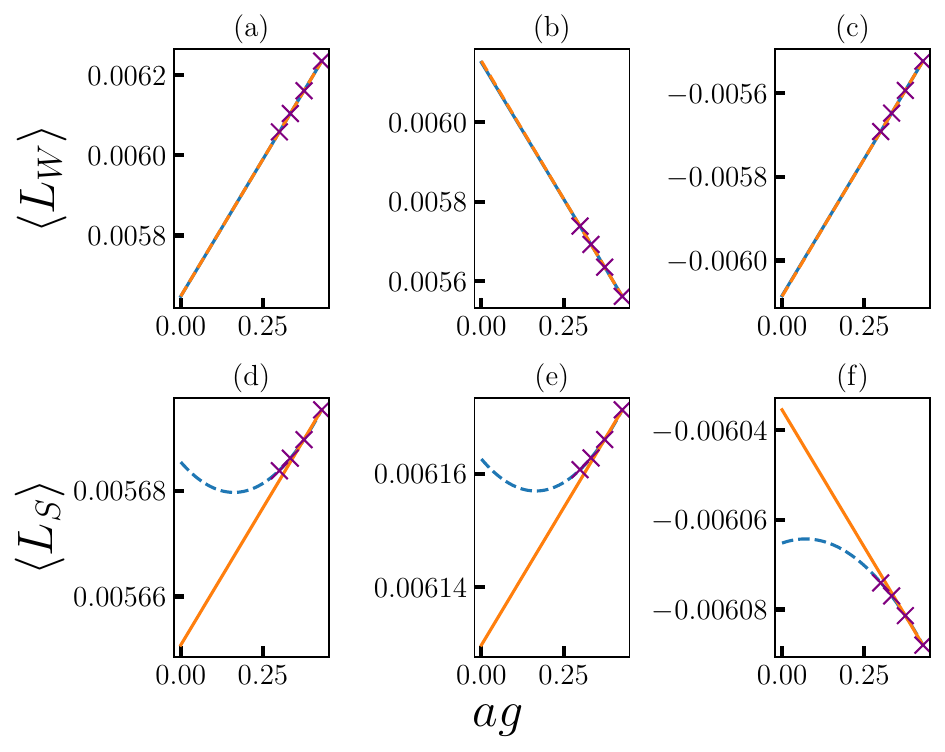}
    \caption{Electric field density $\langle L_{W,S} \rangle$ against $ag$ to extrapolate to the continuum limit with fixed volume $N/\sqrt{x} = 30$, fixed $m_\text{r}/g = 0.01$, using $N = 70, 80, 90, 100$. The continuous orange line is a linear fit and the dashed blue line is a second order polynomial fit. $(a-c)$ is showing Wilson fermions for $l_0 = 0.1, 0.4, 0.6$ respectively and $(d-f)$ the same for staggered fermions with the same $x$ as Wilson fermions and the electric field density method for the mass shift~\cite{angelides}. The error bars emanate from the variational algorithm to compute the relevant ground states and the extrapolation in bond dimension. They are much smaller than the markers and thus, are not visible.}
    \label{efd_vs_ag_extrapolation}
\end{figure}

\begin{figure}[htp!]
    \centering
    \includegraphics[width=\linewidth]{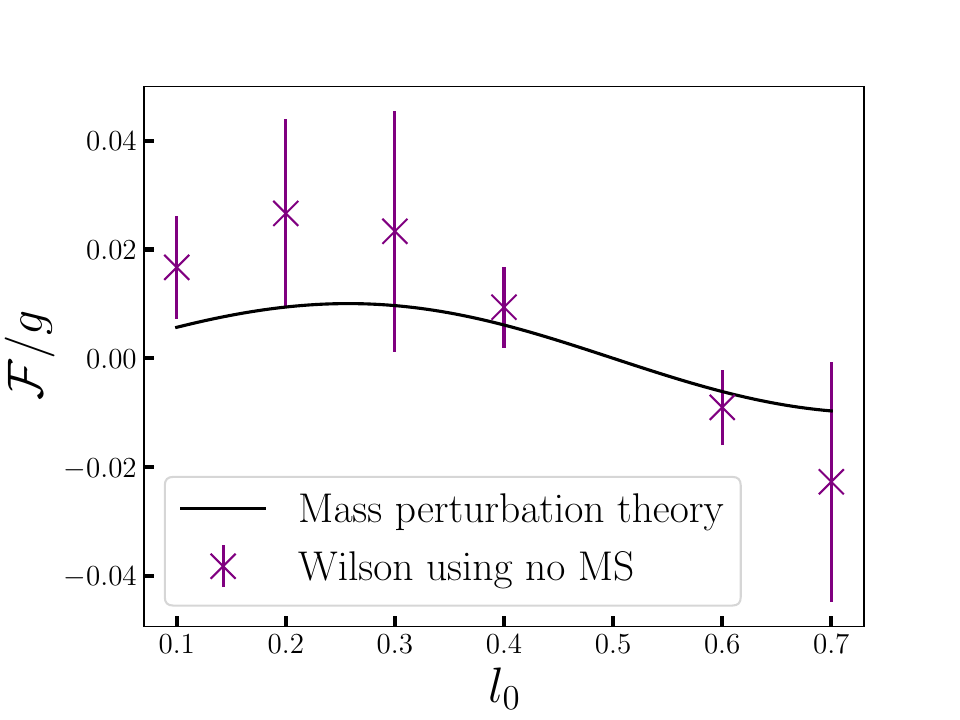}
    \caption{Electric field density $\mathcal{F}/g$ against $l_0$ for Wilson fermions using no mass shift at $m_\text{r}/g = m_\text{lat}/g = 0.01$. The MSE is calculated as in Fig.~\ref{compare_efd_vs_l_0} and found here to be $7.939\cdot10^{-4}$. The error bars emanate from the errors in the variational algorithm to compute the relevant ground states, the extrapolation in bond dimension and in lattice spacing.}
    \label{compare_efd_vs_l_0_Wilson_no_MS}
\end{figure}

After calculating the mass shift, we proceed with the calculation of the electric field density for a fixed renormalized mass of $m_\text{r}/g = 0.01$ and for different values of $l_0$. This procedure enables the comparison of our continuum extrapolation quantities with the theoretical results from mass perturbation theory~\cite{adam}. We use $N$ from 70 to 100 in steps of 10, which in turn determines the $ag = 1/\sqrt{x}$ values through the fixed volume. After bond dimension extrapolations of electric field density, we extrapolate this observable to $ag = 0$ as shown in Fig.~\ref{efd_vs_ag_extrapolation}. For the Wilson case, we take the extrapolated value of the linear fit and assign an approximated error to it which is taken to be the difference between that value and the extrapolated value of a quadratic fit. For the staggered case, it was observed that second-order polynomial fits were more suitable and that they converge faster to the continuum; therefore we keep the extrapolated value of the quadratic fit and again approximate the error as in the Wilson case. To avoid boundary effects, the electric field density for Wilson fermions is measured using the middle link as shown in Eq.~\eqref{L_middle}, and for staggered fermions, the four middle links are measured to also reduce the staggering effect.

\section{Lattice effects on the phase transition\label{prediction_for_transition}}

Here we derive an analytical expression for the location of the first-order phase transition of the lattice Schwinger model focusing on the case of $m/g >> 1$, i.e.\ the limit where the kinetic term is negligible.

We start by mapping the electric field energy to its lattice version. The electric field operator can also be made dimensionless using the coupling $g$ between the fermions and gauge fields, i.e. $E \to gL_n$. The mapping from a continuum theory electric field energy to the corresponding lattice version is
\begin{equation}
    \int_{-\infty}^{\infty} \frac{E^2}{2} dx \to a \sum_{n=0}^{N-2} \frac{g^2L_n^2}{2}.
\end{equation} 

Our derivation will not depend on the specific lattice fermion formulation, hence the result can be seen as general and applying to both Wilson and staggered formulations. The key ingredient is the fact at the point of the first-order phase transition, which we label with $l_0^*$, the ground state is two-fold degenerate, as already described in Sec.~\ref{Continuum_Schwinger_Model}.

On the left of the phase transition ($l_0 < l_0^*$), the ground state has no charges present and the total electric field is $l_0$ on each link. Hence, the total energy $E_{l_0 < l_0^*}$ is just the electric field energy 
\begin{equation}
\label{appendix_energy_equation_for_phase_transition_1}
    E_{l_0 < l_0^*} = a(N-1)g^2\frac{l_0^2}{2},
\end{equation}
where we have used that on a lattice with open boundaries we have $N-1$ links. On the right of the phase transition ($l_0 > l_0^*$), the ground state has a negative charge at the left edge of the system and a positive one at the right edge, connected by an electric flux string lowering total electric field by one unit. This results in a total electric field of $l_0-1$ on each link, and the total energy consists of the electric field energy plus a mass energy of $2m_r$ from the two charges
\begin{equation}
\label{appendix_energy_equation_for_phase_transition_2}
    E_{l_0 > l_0^*} = a(N-1)g^2\frac{(l_0-1)^2}{2} + 2m_r.
\end{equation}
These two energies become equal at $l_0 = l_0^*$, and solving for $l_0^*$ by equating Eq.~\eqref{appendix_energy_equation_for_phase_transition_1} with Eq.~\eqref{appendix_energy_equation_for_phase_transition_2} we find
\begin{equation}
\label{appendix_equation_for_l_0_star}
    l_0^* = 2\frac{m_\text{r}}{g}\frac{\sqrt{x}}{N-1} + \frac{1}{2}.
\end{equation}
In the limit of large $N$, Eq.~\eqref{appendix_equation_for_l_0_star} becomes
\begin{equation}
    l_0^* \approx 2\frac{m_\text{r}}{g}\frac{\sqrt{x}}{N} + \frac{1}{2},
\end{equation}
where $\frac{\sqrt{x}}{N}$ is the dimensionless inverse volume. Hence, in the limit of infinite volume, we obtain the continuum prediction $l_0^* = 1/2$ as expected. If we now take again Eq.~\eqref{appendix_equation_for_l_0_star} and insert for the volume $N/\sqrt{x} = 30$, which we are using in Sec.~\ref{quantum_hardware_results} and Sec.~\ref{results_with_tensor_networks}, and express the renormalized mass in terms of the lattice mass and a mass shift (MS) according to Eq.~\eqref{mass_shift_equation}, we arrive at the result
\begin{equation}
\label{prediction_for_transition_equation}
    l_0^* = \frac{1}{15}\left(\frac{m_\text{lat}}{g} + \text{MS}\right)\frac{1}{1-ag} + \frac{1}{2}.
\end{equation}
The mass shift decreases with decreasing $ag$~\cite{angelides, chiral_dempsey_staggered}. Further, the factor $1/(1-ag)$ decreases as $ag \to 0$, which leads to an overall decrease of the first term and Eq.~\eqref{prediction_for_transition_equation} tends towards the continuum limit. Hence, we see that the transition point on finite lattices $l_0^*$ is generally shifted towards values larger than $1/2$ and gets closer to the continuum theory prediction for decreasing lattice spacing, as observed in Sec.~\ref{quantum_hardware_results} and supported by Fig.~\ref{efd_vs_l_0_real_hardware_comparison_mass_10}.

\end{appendix}

\begin{table*}
\centering
\resizebox{\linewidth}{!}{
\begin{tabular}{|c|c|c|c|c|c|c|}
\hline
$l_0$                                                                                           & 0.1            & 0.2            & 0.3            & 0.4            & 0.6            & 0.7            \\ \hline
Wilson                                                                                          & 4.45001467e-05 & 1.00338853e-05 & 5.58001106e-05 & 3.04735487e-05 & 3.64867975e-05 & 2.17138610e-05 \\ \hline
\begin{tabular}[c]{@{}c@{}}Staggered \\  ($\text{MS}_t$, same num. qubits as Wilson)\}\end{tabular} & 0.00012783     & 0.00016672     & 0.00022494     & 0.0003207      & 0.00045197     & 0.00046271     \\ \hline
\begin{tabular}[c]{@{}c@{}}Staggered\\ ($\text{MS}_{L}$, same x as Wilson)\end{tabular}                  & 6.97181950e-06 & 4.86364083e-06 & 1.14145910e-05 & 4.21571198e-05 & 5.52293937e-05 & 3.33572489e-05 \\ \hline
\begin{tabular}[c]{@{}c@{}}Staggered\\ ($\text{MS}_t$, same x as Wilson)\end{tabular}               & 2.91080596e-04 & 2.63562909e-04 & 8.61951831e-05 & 1.03764598e-04 & 6.79235186e-04 & 6.37947680e-04 \\ \hline
\begin{tabular}[c]{@{}c@{}}Staggered\\ ($\text{MS}_{L}$, same num. qubits as Wilson)\end{tabular}        & 2.05654489e-05 & 1.75062880e-05 & 1.54209990e-05 & 1.40152738e-04 & 5.40892433e-05 & 5.13089222e-05 \\ \hline
\end{tabular}}
\caption{This table corresponds to the absolute error with respect to the results from mass perturbation theory for the data presented in Fig.~\ref{compare_efd_vs_l_0}. For staggered fermions that use the theoretically predicted mass shift, $\text{MS}_t$, the error tends to grow with $l_0$ due to the fact that the $\text{MS}_t$ does not account for an $l_0$ dependence.}
\label{compare_efd_vs_l_0_table}
\end{table*}

\FloatBarrier
\bibliography{References}

\begin{acknowledgments}

    This work is partly funded by: the European Union’s Horizon 2020 Research and Innovation Programme under the Marie Sklodowska-Curie COFUND scheme with grant agreement no.\ 101034267, the European Union's Horizon Europe Framework Programme (HORIZON) under the ERA Chair scheme with grant agreement no.\ 101087126 and the Ministry of Science, Research and Culture of the State of Brandenburg within the Centre for Quantum Technologies and Applications (CQTA). P.N.\ acknowledges financial support from the Cyprus Research and Innovation Foundation under projects ``Future-proofing Scientific Applications for the Supercomputers of Tomorrow (FAST)'', contract no.\ COMPLEMENTARY/0916/0048, and ``Quantum Computing for Lattice Gauge Theories (QC4LGT)'', contract no.\ EXCELLENCE/0421/0019. A.C.\ is supported in part by the Helmholtz Association
    —“Innopool Project Variational Quantum Computer
    Simulations (VQCS).” The authors also thank the QC4HEP Working Group for discussions. IBM, the IBM logo, and ibm.com are trademarks of International
    Business Machines Corp., registered in many jurisdictions
    worldwide. Other product and service names might be trademarks of IBM or other companies.
    \begin{figure}[H]
        \centering
        \includegraphics[width = 0.08\textwidth]{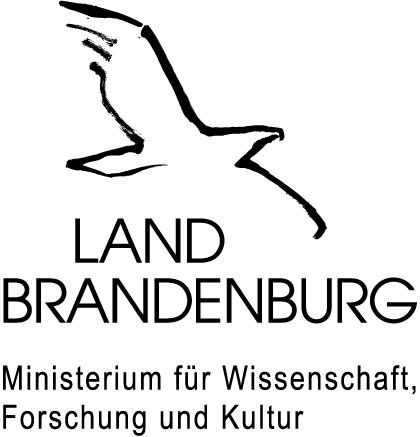}
    \end{figure}
\end{acknowledgments}

\section*{AUTHOR CONTRIBUTIONS}
T.A. drafted the paper, led the calculations for Wilson fermions and performed the MPS computations. P.N. led the calculations for staggered fermions. D.W. advised on hardware experiments with error mitigation. Everyone has contributed to providing directions and ideas and writing the paper.  

\section*{COMPETING INTERESTS}

All authors declare no financial or non-financial competing interests.

\end{document}